\documentclass[11pt]{article}
\usepackage{graphicx}
\usepackage{epsfig}
\usepackage{epstopdf}
\usepackage{amsmath}
\usepackage{amsfonts}
\usepackage{amssymb}

\usepackage[english]{babel}

\usepackage{hyperref}

\newcommand{\eq}{\begin{equation}}
\newcommand{\beq}{\begin{equation}}
\newcommand{\eeq}{\end{equation}}
\newcommand{\eqa}{\begin{eqnarray}}
\newcommand{\eeqa}{\end{eqnarray}}
\newcommand{\beqa}{\begin{eqnarray}}
\newcommand{\bea}{\begin{eqnarray}}
\newcommand{\eea}{\end{eqnarray}}

\newcommand{\mc}[1]{\mathcal{#1}}
\newcommand{\dd}{{\textrm{d}}}
\newcommand{\w}{{\wedge}}

\newcommand{\La}{{\Lambda}}

\newcommand{\p}{{\partial}}

\newcommand{\be}{{\beta}}

\newcommand{\ep}{{\epsilon}}

\newcommand{\si}{{\sigma}}

\newcommand{\om}{{\omega}}
\newcommand{\Om}{{\Omega}}

\newcommand{\lp}{\left(}
\newcommand{\rp}{\right)}

\newcommand{\lb}{\left[}
\newcommand{\rb}{\right]}

\newcommand{\half}{\frac{1}{2}}

\newcommand{\BE}{\begin{equation}}
\newcommand{\EE}{\end{equation}}
\newcommand{\BC}{\begin{center}}
\newcommand{\EC}{\end{center}}
\newcommand{\BF}{\begin{figure}}
\newcommand{\EF}{\end{figure}}

\begin{document}

\setlength{\unitlength}{1mm}

\thispagestyle{empty}
\begin{flushright}
\small \tt
\begin{tabular}{l}
CPHT-RR004.0212\\
LPT-ORSAY 12-15
\end{tabular}
\end{flushright}
\vspace*{1.cm}

\begin{center}
{\bf \LARGE Shaping black holes with free fields}\\

\vspace*{1.5cm}

{\bf Yannis Bardoux,}$^{1}\,$
{\bf Marco M.~Caldarelli,}$^{2,1}\,$
{\bf Christos Charmousis,}$^{1,3}\,$

\vspace*{0.5cm}

{\it $^1$\,Laboratoire de Physique Th\'eorique, Univ. Paris-Sud,\\
 CNRS UMR 8627,
 F-91405 Orsay, France}\\[.3em]

{\it $^2$\,Centre de Physique Th\'eorique, Ecole Polytechnique,\\
 CNRS UMR 7644,
 F-91128 Palaiseau, France}\\[.3em]

{\it $^3$\,Laboratoire de Math\'ematiques et Physique Th\'eorique (LMPT), Univ. Tours,\\
UFR Sciences et Techniques,\\
Parc de Grandmont, F-37200 Tours, France}\\[.3em]

\vspace*{0.3cm} {\small\tt
yannis.bardoux@th.u-psud.fr,
marco.caldarelli@th.u-psud.fr,
christos.charmousis@th.u-psud.fr
}

\vspace*{.3cm}

\vspace{.8cm} {\bf ABSTRACT}
\end{center}

\noindent

Starting from a metric Ansatz permitting a weak version of Birkhoff's theorem we find static black hole solutions including matter in the form of free scalar and $p$-form fields, with and without a cosmological constant $\La$. Single $p$-form matter fields permit multiple possibilities, including dyonic solutions, self-dual instantons and metrics with Einstein-K\"alher horizons. The inclusion of multiple $p$-forms on the other hand, arranged in a homogeneous fashion with respect to the horizon geometry, permits the construction of higher dimensional dyonic $p$-form black holes and four dimensional axionic black holes with flat horizons, when $\La<0$. It is found that axionic fields regularize black hole solutions in the sense, for example, of permitting regular -- rather than singular -- small mass Reissner-Nordstrom type black holes. Their cosmic string and Vaidya versions are also obtained.


\vfill \setcounter{page}{0} \setcounter{footnote}{0}
\newpage

\tableofcontents



\section{Introduction and setup}

Theorems regarding black hole uniqueness \cite{unique} for static and stationary spacetimes respectively, led Wheeler to his famous conjecture stating that black holes have no hair. The conjecture states, that apart from charges measured at infinity by a far away observer, no additional degrees of freedom can  describe the black hole geometry (for a review see \cite{bek1}). We know that in a four dimensional  Einstein-Maxwell theory and for a stationary and asymptotically flat spacetime the only possible parameters are mass, angular momentum, electric and magnetic monopole charge. The conjecture questions the extension  of this fact to more generic theories and weaker hypotheses. The physical idea behind this conjecture is that by the time a black hole relaxes into a stationary state it will have either expelled or eaten up all physical degrees of freedom in its vicinity, leaving only those corresponding to far-away conserved charges as measured via a Gauss law. This statement -- if true -- has important physical consequences. For example, a neutron star described amongst other things by lepton or baryon number would shave its hair if it were to collapse to a black hole. A black hole, according to the conjecture, is a rather blunt and bald gravitational object having specific charges and not allowing additional parameters -- primary hair -- which are not associated to a conserved charge, or, secondary hair of no additional parameters but non-trivial fields interacting with the black hole spacetime. This again means that black holes which can be loosely interpreted as gravitational solitons (asymptotically well behaved and finite energy objects) would not acquire excited states provided as additional hair.  The real  question underlying this conjecture, is under  which hypotheses is the conjecture actually valid, or, in a weaker version, when is it not. 

Multiple ways were found of circumventing this conjecture by evading one of the hypotheses of the black hole theorems or by including some non-trivial matter fields and couplings in between them. Changing for example the asymptotic properties of the black hole, by implementing a cosmological constant, or allowing for non-trivial  topology will introduce long distance hair such as those of an abelian Higgs vortex \cite{ruth} ending or piercing  a black hole. Or again, one could have non-abelian gauge fields, providing colour for  black holes \cite{volkov} as primary hair or in some cases black holes embedded in magnetic monopoles \cite{weinberg}. 
Interestingly, upon coupling conformally a scalar field to curvature we can have compelling solutions.
The BBMB \cite{bbmb} static and spherically symmetric solution  is the closest one can get to a massless scalar-tensor black hole\footnote{Although the geometry is regular the scalar field explodes on the black hole horizon and the black hole interpretation is not clear \cite{bek1}.}. The scalar has no associated charge and the hair is secondary, emerging from the particular conformal coupling of the scalar field. The spacetime geometry  is that of the extremal Reissner-Nordstrom. In fact, as shown in \cite{Xanthopoulos:1991mx} any departure from extremality leads to a singular geometry.  If one now allows for a cosmological constant the exploding scalar is pushed within the horizon \cite{mtz} and the solution is a genuine black hole with  secondary hair. Generically however, and in favour of the conjecture if one asks only for stable black hole solutions, many of the above spacetimes do not pass the test as they are  perturbatively unstable\footnote{P. McFadden et al.~claimed the inverse about the BBMB solution \cite{McFadden:2004ni} and to our knowledge the question is still unsettled.} \cite{unstable}. Exceptions include the abelian Higgs vortex  and the skyrmion black hole \cite{skyrme} protected by topological charge. 

But what is the situation concerning the dressing of $D$-dimensional black holes with free $p$-form fields beyond the case of electromagnetism? This includes the case of cosmological constant, multiple scalar fields, three-forms, spacetime filling forms  and so on.  For a start we expect that $p$-forms (at least with $p>1$), if allowed to act as monopole charge, will give rise to some conserved charge at infinity. Hence they will not be classified as hair, at least in the terms given above. They will however dress the horizon of the black hole often rendering novel horizons or singularities, sometimes completely changing the properties of the solution. A typical example is that of the Reissner-Nordstrom geometry where the charge, when lower than mass, creates an inner horizon completely changing the nature of the central curvature singularity and, for a sufficiently high charge, a naked singularity. Furthermore the inclusion of $p$-form matter will also, in some cases, change the asymptotic properties of the black hole spacetime. In fact we will see that the $p$-forms can act as external fields much like the Melvin  homogeneous magnetic field dressing the Schwarzschild black hole \cite{ernst}.  Recently, Emparan et al.~\cite{Emparan:2010ni} argued on the non-existence of static black holes dressed by a $p$-form field (with $(D+1)/2\leq p \leq D-1 $) assuming the presence of a static regular horizon and asymptotic flatness. Their argument, later generalized by Shiromizu et al.~\cite{Shiromizu:2011he} to include any $p\geq3$ form field strength under the extra assumption of spherical topology, does not exclude $p$-forms carried by horizons with non-spherical topology, nor forbids $p$-form fields
with distorted asymptotics. In fact, Emparan gave an explicit solution of a three-form dipole as primary hair for  the black ring \cite{Emparan:2004wy}. One also expects, much like Melvin spacetimes, that the fall-off properties of the solution will change for higher $p$-form fields, thus inevitably changing the asymptotic properties of the black hole. In other words the asymptotic flatness hypothesis is maybe not adapted to the case of  $p$-form  black holes.  In this paper we will study the problem of $p$-form dressing of static black holes in quite some generality and show a plethora of novel solutions. In particular, we will exhibit the first static four dimensional black hole with non trivial three-form (axionic) charge. We will show how axions regularize electrically charged solutions. We will also see how scalar fields dressing a toroidal horizon, while breaking horizon space symmetries, can give AdS black hole geometries with flat horizon, but with a lapse function of the form usually associated with hyperbolic black holes. The scalar fields act as vacuum energy for the horizon geometry creating an effective cosmological constant as in cosmological self-tuning scenarios \cite{fab4}. Finally, some of these solutions sporting a negative cosmological constant can have applications to AdS/CFT and condensed matter applications.

We will complete this introductory section, by presenting our metric Ansatz and reviewing the vacuum black holes that it contains. Next, in Section~\ref{sec::fields} we will introduce the matter fields and adapt them to the spacetime geometry. Black holes dressed by a single matter field will be discussed in Section~\ref{sec::single}, while Section~\ref{sec::kaehler} will treat the special case of Einstein-K\"ahler horizons. When the curvature of the horizon is positive, black holes presented in those sections exist for any cosmological constant $\La$, otherwise they are AdS black holes. We will then discuss in Section~\ref{sec::multiple} the black holes with multiple $p$-form matter fields. This possibility relies on the presence of a negative cosmological constant, and the particular case of axionic black holes will be analyzed in details in Section~\ref{sec::axion}. The thermodynamics of these black holes is left for Section~\ref{sec::thermo}, while concluding remarks will be provided in Section~\ref{sec::conclusion}.

\paragraph{An Ansatz for static black holes.}
It is well-known that in four spacetime dimensions, in presence of a negative cosmological constant, the topological censorship theorems can be evaded and asymptotically locally anti-de~Sitter black holes with flat or hyperbolic horizons can be constructed. Upon compactification, arbitrary topology event horizons can be constructed \cite{topological,Vanzo:1997gw}, at the cost of having a non-trivial topology at spatial infinity \cite{Galloway:1999bp}. These black holes, usually referred to as topological black holes, are easily generalized to higher dimensions \cite{Birmingham:1998nr}, and stem from the existence of solutions with extended event horizons, whose intrinsic geometry is flat or hyperbolic in four dimensions, or more generally an Einstein manifold when $D>4$.

A convenient and quite general starting point for the study of such metrics is the following warped ansatz,
\eq \label{metric0}
ds^2=-2e^{2\nu(u,v)}B(u,v)^{-\frac{n}{n+1}}du\,dv+B(u,v)^{\frac{2}{n+1}} \sigma_{ij}(y) dy^idy^j,
\eeq
where we have used light-cone coordinates $(u,v)$
and have defined $D=n+3$ for convenience, with $n\geq1$. The metric is time-dependent, parameterized by two independent functions $\nu(u,v)$ and $B(u,v)$, while $\si_{ij}(y)$ is an arbitrary Riemannian signature metric of some smooth $(n+1)$-dimensional transverse manifold $\mathcal H$. This metric includes all static metrics, and also all topological black holes. As discussed recently in the context of Einstein Gauss-Bonnet gravity in vacuum \cite{Bogdanos:2009pc} and in the presence of $p$-forms \cite{yannisteo}, such a class of metrics obeys a weak version of Birkhoff's staticity theorem stating the existence of a local timelike Killing vector field. The four dimensional version of (\ref{metric0}) (without a negative cosmological constant) gives us the usual uniqueness theorem of Birkhoff: the only asymptotically flat spherically symmetric solution of general relativity is given by the Schwarzschild geometry{\footnote{In this paper when we refer to Birkhoff's theorem we will always refer to its generalized weaker version.}} 
The generalization of the Schwarzschild black hole to Einstein Gauss-Bonnet gravity was found by Boulware and Deser \cite{Boulware:1985wk}, whereas Wiltshire \cite{Wiltshire:1985us}, was first to demonstrate the relevant Birkhoff theorem.
This result is true even in the presence of matter, as long as the energy-momentum tensor obeys the condition (see also \cite{Jacobson:2007tj}),
\beq
\label{mattercond}
T_{uu}=T_{vv}=0.
\eeq
Using the $uu$ and $vv$ components of Einstein's equations one reduces (\ref{metric0}) without further hypothesis to the well known metric 
\eq
ds^2=-V(r)dt^2+\frac{dr^2}{V(r)}+r^2\si_{ij}(y)dy^idy^j,
\label{metric}\eeq
where $V(r)$ and $\si_{ij}(y)$ are to be determined by the remaining field equations.

The Ricci tensor associated to this geometry is
\eqa
&&R_{tt}=-V^2R_{rr}=\frac12VV''+\frac{n+1}{2r}VV'=\frac V{2r^{n+1}}\lp r^{n+1}V'\rp',\label{Rtt}\\
&&R_{ij}=\mc R_{ij}-(rV'+nV)\si_{ij}=\mc R_{ij}-\frac1{r^{n-1}}\lp r^nV\rp'\si_{ij},
\eeqa
with $\mc R_{ij}$ the Ricci tensor of the manifold $\mc H$, obtained from its intrinsic metric $\si_{ij}$.
Given a negative cosmological constant
\eq
\Lambda=-\frac{(D-1)(D-2)}{2\ell^2}=-\frac{(n+2)(n+1)}{2\ell^2},
\eeq
and no extra matter fields, Einstein's equations read
\eq
R_{\mu\nu}=-\frac{n+2}{\ell^2}g_{\mu\nu}.
\eeq
By \eqref{Rtt}, the $tt$ and $rr$ components of these equations are proportional to each other, and solved by the potential
\eq
V(r)=\kappa-\frac{r_0^n}{r^n}+\frac{r^2}{\ell^2},
\label{vacuum}\eeq
with $\kappa$ and $r_0$ two integration constants. Then, the $ij$ components reduce to \cite{Vanzo:1997gw,Birmingham:1998nr}
\eq
\mc R_{ij}=n\kappa\si_{ij},
\label{Einstein manifold}\eeq
and require $\mc H$ to be an Einstein space, with curvature set by $\kappa$.
These solutions possess an event horizon as long as $r_0$ is large enough, and describe the geometry of topological black holes \cite{topological,Vanzo:1997gw,Birmingham:1998nr}. When $\kappa\leq0$, the negative cosmological constant is crucial to have an event horizon hiding the central singularity.

On the other hand, if the Einstein manifold $\mc H$ is of positive curvature ($\kappa>1$), the solution survives as a black hole when the cosmological constant is continued to $\La=0$ or $\La>0$. In the former case, when $\mc H=\mc S^{n+1}$ with the usual unit round metric and
\eq
V(r)=\kappa-\frac{r_0^n}{r^n},
\eeq
we recover the generalization of the Schwarzschild black hole to higher dimensions obtained by Tangherlini \cite{Tangherlini:1963bw}; for any other compact Einstein manifold satisfying \eqref{Einstein manifold} with $\kappa>0$, it yields a generalized Schwarzschild-Tangherlini black holes \cite{gibbonshartnoll}.
For example, following \cite{Gibbons:2002bh},
one can keep $\mc H$ of spherical $\mc S^{n+1}$ topology, but endowed with an inhomogeneous B\"ohm metric \cite{bohm}, at least for $4\leq n\leq8$. Another possibility, with different topology, is to take $\mc H$ to be a product of spheres carrying a $\kappa>0$ Einstein metric. The same construction works with positive cosmological constant, giving de~Sitter black holes with
\eq
V(r)=\kappa-\frac{r_0^n}{r^n}-\frac{r^2}{\ell^2}.
\eeq
The only difference comes from the presence of a cosmological horizon, that restricts the range of the $r_0$ parameter for which the de~Sitter universe contains a  black hole. The reader should keep in mind that, whenever we discuss AdS black holes with $\kappa>0$, it is possible to continue the solution to $\La\geq0$ and obtain a matching asymptotically locally flat/de~Sitter black hole, with same $\mc H$ and matter fields. This is is true in particular for the $\kappa>0$ solutions displayed in Section~\ref{sec::single} and Section~\ref{sec::kaehler}, although we shall not emphasize this possibility each time for concision's sake.

The question we shall investigate in the rest of this article is, to what extent matter fields can dress these solutions, obtaining static black holes with metric of the form \eqref{metric}.

\paragraph{An ansatz for the stress tensor of the external matter fields, and its effect on the geometry.}
In this article, we are interested in geometries that can be kept in the form \eqref{metric}, even as matter fields are included. As explained above, this means that the total stress tensor $T_{\mu\nu}$ must satisfy \eqref{mattercond} in light-cone coordinates, thus admitting the aforementioned weak version of Birkhoff's theorem. In the $tr$ coordinates, this is rephrased into the matter tensor satisfying conditions $T_{tr}=0$ and $T_{tt}+V^2T_{rr}=0$. It is well known that the energy-momentum tensor of a Maxwell field satisfies this condition whereas, as we will see, radially or time dependent scalars do not. In fact, radially dependent scalars lead to singular solutions, whereas the matter contraint we impose here, filters these away and points towards black hole geometries. In this sense, the matter constraint we impose is physically motivated and a sensible regularity constraint. Moreover, we do not want to bring into play any additional privileged vectors or tensors, other than those coming from the particular foliation introduced by the metric ansatz. We will refer hereafter to this property as `isotropy' of the stress tensor. Finally we require that the stress tensor cannot be used to distinguish different points on $\mc H$, that is, the total distribution of stress tensor is `homogeneous'\footnote{This is not to be confused with the (in-)homogenity of the metric $\si_{ij}$ carried by $\mc H$.}. Under those assumptions, it follows that the stress tensor is fully determined by two functions of $r$ only -- that we name $\ep(r)$ and $P(r)$ -- and assumes the general diagonal form
\eq
T_{\mu\nu}=\frac{1}{16\pi G r^{n+1}}\lp
\begin{array}{ccc}
V(r)\ep(r)&&\\
&-\ep(r)/V(r)&\\
&&r^{2}P(r)\si_{ij}
\end{array}
\rp.
\label{st}\eeq
With such a source, Einstein's equations in presence of the cosmological constant $\Lambda$,
\eq
G_{\mu\nu}+\Lambda g_{\mu\nu} = 8\pi G T_{\mu\nu}
\eeq
reduce to the system,
\eq
\lp r^{n+1}V'\rp'=\frac{2(n+2)}{\ell^2}r^{n+1}+\frac{n-1}{n+1}\ep(r)+P(r),
\label{eq1}\eeq
\eq
\mc R_{ij}=\frac1{r^{n-1}}\lb{\lp r^nV\rp'}-\frac{n+2}{\ell^2}{r^{n+1}}+\frac{\ep(r)}{n+1}\rb\si_{ij}.
\eeq
Since both the metric $\si_{ij}$ and the Ricci tensor $\mc R_{ij}$ of $\mc H$ depend only on the transverse $y^i$ coordinates, the proportionality factor must be a constant, that we dub $n\kappa$,
\eq
{\lp r^nV\rp'}-\frac{n+2}{\ell^2}{r^{n+1}}+\frac{\ep(r)}{n+1}=n\kappa{r^{n-1}},
\label{eqV}\eeq
and therefore $\mc H$ is an Einstein manifold, satisfying \eqref{Einstein manifold}.
Equation \eqref{eqV} can be integrated to obtain the lapse function,
\eq
V(r)=\kappa-\frac{r_0^n}{r^n}+\frac{r^2}{\ell^2}-\frac1{r^n}\int\frac{\ep(r)}{n+1}dr,
\label{V}\eeq
with $r_0$ an integration constant of the dimensions of a length, and the remaining equation \eqref{eq1}, equivalent to the conservation of the stress tensor, is
\eq
\ep'(r)+\frac{n+1}rP(r)=0.
\label{conservation}\eeq
As long as this equation holds, the metric \eqref{metric} with lapse function \eqref{V} and $\mc H$ an Einstein manifold \eqref{Einstein manifold} solves Einstein's equations sourced by \eqref{st}. In the rest of the paper, we will show how stress tensors of this form can be obtained with various combinations of free fields. We will then conclude by studying simple properties they enjoy.

\paragraph{Notation and conventions:} in what follows, we shall consider static $D$-dimensional spacetimes with metric $g_{\mu\nu}$, and define $n=D-3$ for convenience. Spacetime indices are denoted by Greek letters $\mu,\nu,\ldots$. Constant time and radial coordinate sections of dimension $n+1$ are denoted $\mc H$ and have induced metric $\si_{ij}$. Latin indices $i,j,\ldots$ are tangent to these submanifolds. Latin indices $a,b,\ldots$ are used to label single factors of $\mc H$ when the latter takes the form of a direct product $\mc H=\mc H^{(a_1)}\times\cdots\times\mc H^{(a_m)}$, with the induced metric on the factors given by $\si^{(a_k)}_{ij}$. The volume forms of the spacetime and of the sections $\mc H$ are given by $\ep$ and $\hat\ep$ respectively. Finally, we will consider form fields strength $H_{[p]}=\dd B_{[p-1]}$ of rank $p$ in the matter sector, dropping the index $[p]$ when it is obvious.

\section{Free scalar and $p$-form fields}\label{sec::fields}

To minimally couple a free $p$-form field strength $H_{[p]}=\dd B_{[p-1]}$ to the gravitational field, we start with the Einstein-Hilbert action
\eq
S_{0}=\frac{1}{16\pi G}\int d^Dx\sqrt{-g}\,\lp R-2\Lambda\rp,
\label{action}\eeq
and add a matter term of the form,
\eq
S_M=-\frac1{16\pi G}\int d^Dx\sqrt{-g}\,\frac{1}{2p!}H_{[p]}^2.
\label{SM}\eeq
The equations of motion and the Bianchi identity for $H_{[p]}$ read
\eq
\nabla_{\mu}H^{\mu\nu_1\ldots\nu_{p-1}}=0,\qquad
\nabla_{[\mu}H_{\nu_1\ldots\nu_p]}=0,
\label{eqH}\eeq
and its stress tensor is given by
\eq
T_{\mu\nu}=\frac1{16\pi G(p-1)!}\lp H_{\mu\rho_1\ldots\rho_{p-1}}H_\nu{}^{\rho_1\ldots\rho_{p-1}}-\frac1{2p}H^2g_{\mu\nu}\rp.
\eeq
Imposing the constraints \eqref{mattercond} on such a stress tensor, we observe that the relation $T_{tt}+V^2T_{rr}=0$ reduces to a sum of squares. This implies in turn that the components of the field with all but one leg on the $\mc H$ directions have to vanish,
\eq
H_{ti_1\ldots i_{p-1}}=H_{ri_1\ldots i_{p-1}}=0.
\eeq
Note that for free scalars ($p=1$), this condition is saying that the fields are independent of $r$ and $t$ coordinates. This is not too surprising, scalar fields generically excite radial breather modes breaking Birkhoff's theorem and simple counterexamples are known (see for example \cite{liouville}). Accordingly, here we see that when they depend only on the horizon coordinates they do not break Birkhoff's theorem.  
Then, equations \eqref{eqH} are solved by
\eq
H_{tri_1\ldots i_{p-2}}=\frac{1}{r^{n-2p+5}}{\mc E}_{i_1\ldots i_{p-2}}(y),\qquad
H_{i_1\ldots i_{p}}={\mc B}_{i_1\ldots i_{p}}(y).
\label{H}\eeq
Using the internal metric $\si_{ij}$ to raise and lower the indices of the tensors $\mc E$ and $\mc B$, we have
\eq
H^{tri_1\ldots i_{p-2}}=-\frac{1}{r^{n+1}}{\mc E}^{i_1\ldots i_{p-2}}(y),\qquad
H^{i_1\ldots i_{p}}=\frac1{r^{2p}}{\mc B}^{i_1\ldots i_{p}}(y).
\label{Hup}\eeq
Here, $\mc E$ and $\mc B$ are rank $p-2$ and rank $p$ form fields on $\mc H$ respectively, such that
\eq
\p_{i_1}\lp\sqrt\si{\mc E}^{i_1\ldots i_{p-2}}\rp=0,\qquad
\p_{[j}{\mc E}_{i_1\ldots i_{p-2}]}=0,
\label{eqE}\eeq
\eq
\p_{i_1}\lp\sqrt\si{\mc B}^{i_1\ldots i_{p}}\rp=0,\qquad
\p_{[j}{\mc B}_{i_1\ldots i_{p}]}=0.
\label{eqB}\eeq
These harmonic forms define the polarization on $\mc H$ of the electric and magnetic parts of the field $H$ and, as we shall show in Section~\ref{sec::thermo}, they correspond to the conserved charges associated to the field $H$.
The associated stress tensor has components
\eqa
&&\displaystyle T_{tt}=-V^2T_{rr}=\frac V{16\pi G}\lp
\frac{\mc E^2}{2(p-2)!r^{2n-2p+6}}+\frac{\mc B^2}{2p!r^{2p}}
\rp,\\[1ex]
&&\displaystyle T_{ij}=\frac1{16\pi G}\lb
-\frac{1}{(p-3)!r^{2n-2p+4}}\lp\mc E_{ik_1\ldots}\mc E_{j}{}^{k_1\ldots}-\frac1{2(p-2)}\mc E^2\si_{ij}\rp\right.\nonumber\\
&&\displaystyle\qquad\qquad\qquad\qquad\qquad\qquad\left.+\frac1{(p-1)!r^{2p-2}}\lp\mc B_{ik_1\ldots}\mc B_{j}{}^{k_1\ldots}-\frac1{2p}\mc B^2\si_{ij}\rp\rb.
\eeqa
To obtain a stress tensor of the form \eqref{st}, the component $T_{tt}$ cannot depend on the transverse coordinates $y^i$. When $2p\neq n+3$, this implies that $\mc E^2$ and $\mc B^2$ must be constants, but in the $2p=n+3$ case the electric and magnetic terms scale with the same power of $r$, and only the constancy of $\mc B^2+p(p-1)\mc E^2$ follows. Therefore, the invariants $\mc E^2$ and $\mc B^2$ could in principle depend on the coordinates $y^i$ on $\mc H$, as long as these dependencies cancel. This would on the other hand violate our homogeneity hypothesis (or, equivalently, we could construct two vectors $\hat\nabla_i\mc E^2$ and $\hat\nabla_i\mc B^2$ that break isotropy). Hereafter, we shall simply assume that both $\mc E^2$ and $\mc B^2$ are constants, for any $p$. Then, we can read the energy density $\ep(r)$ from $T_{tt}$,
\eq
\ep(r)=
\frac{\mc E^2}{2(p-2)!r^{n-2p+5}}+\frac{\mc B^2}{2p!r^{2p-n-1}},
\label{ep}\eeq
and define a pressure by $P(r)=16\pi G T_{ij}\si^{ij} r^{n-1}/(n+1)$,
\eq
P(r)=\frac1{n+1}\lp
\frac{n-2p+5}{2(p-2)!r^{n-2p+5}}\mc E^2
+\frac{2p-n-1}{2p!r^{2p-n-1}}\mc B^2
\rp . 
\eeq
These quantities satisfy automatically the conservation equation \eqref{conservation}. The last constraint to obtain a stress tensor of the form \eqref{st} comes from the isotropy and homogeneity on $\mc H$, which imposes, when $2p\neq n+3$,
\eq
\mc E_{ik\ldots}\mc E_j{}^{k\ldots}=\frac{\mc E^2}{n+1}\si_{ij},\qquad
\mc B_{ik\ldots}\mc B_j{}^{k\ldots}=\frac{\mc B^2}{n+1}\si_{ij}.
\label{isotropy}\eeq
Again, when the spacetime dimension is even and $2p=n+3$, the $r$-dependence of the electric and the magnetic parts of $T_{ij}$ coincide, and the isotropy constraint is weakened,
\eq
\mc B_{ik\ldots}\mc B_j{}^{k\ldots}-\frac{\mc B^2}{n+1}\si_{ij}=(p-1)(p-2)\lp\mc E_{ik\ldots}\mc E_j{}^{k\ldots}-\frac{\mc E^2}{n+1}\si_{ij}\rp.
\label{dyonicisotropy}\eeq
In this case, we will see that dyonic solutions exist. Observe that one can define a new rank two anti-symmetric tensor for dyonic solutions, as the contraction of the electric and magnetic polarization forms, $\mc A_{ij}=\mc B_{ijk_1\ldots k_{p-2}}\mc E^{k_1\ldots k_{p-2}}$. Potentially, $\mc A_{ij}$ could break isotropy. As we will see, it turns out that electric and magnetic fluxes of dyonic solutions are carried by orthogonal spaces and $\mc A_{ij}$ vanishes, unless $\mc H$ is a direct product of two-dimensional spaces, in which case $\mc A_{ij}$ can be proportional to the volume forms of these two-dimensional spaces, without introducing additional privileged directions.
Finally, note that the trace part of equation \eqref{dyonicisotropy} is automatically verified.

Once two forms $\mc E$ and $\mc B$ solving equations \eqref{eqE}, \eqref{eqB} and \eqref{isotropy} or \eqref{dyonicisotropy} are given, we obtain in this way a solution of the gravitational equations coupled to the $p$-form field strength. We shall not attempt a full classification of the possible solutions, but content ourselves to construct the simplest solutions out of the natural tensors that are available on $\mc H$. If no extra structure is present, the only anti-symmetric tensor of $\mc H$ that can be used is the volume form $\hat\ep$ on $\mc H$, but if the transverse space is a K\"ahler space, we can build the polarization vectors out of the K\"ahler forms too. We will start doing so with one single form field, and then extend the construction in cases where multiple form fields $H_{[p]}^{(i)}$ are available.

\section{Black holes dressed by a single field}\label{sec::single}

When the rank of the form $\mc E$ or $\mc B$ is equal to the dimension of $\mc H$, the isotropy condition \eqref{isotropy} is easily met by taking the corresponding form to be proportional to the volume form on $\mc H$. In addition, the volume form automatically satisfies equations \eqref{eqE} or \eqref{eqB}. Denoting the volume form on $\mc H$ by $\hat\ep_{[n+1]}$, we can have a non-vanishing $\mc E=q_e\hat\ep_{[n+1]}$ when $p=n+3$, giving an electrically charged solution, and a non-vanishing $\mc B=q_m\hat\ep_{[n+1]}$ when $p=n+1$, yielding a magnetically charged solution.

\paragraph{Electric $p=2$ solutions, for any $n$:} this is simply Einstein-Maxwell theory with $\Lambda<0$, and the electric field, being $\mc E$ a $0$-form, can be turned on for any $n$ without breaking isotropy. Taking $\mc E=q_e$ one can directly apply the previous results to obtain the Reissner-Nordstrom-AdS solution,
\eq
V(r)=\kappa-\frac{r_0^n}{r^n}+\frac{r^2}{\ell^2}+\frac{q_e^2}{2n(n+1)r^{2n}},
\qquad
H_{tr}=\frac{q_e}{r^{n+1}}.
\eeq

\paragraph{Magnetic $p=n+1=D-2$ solution:} this is dual to an electric two-form, and the solution is the dual of the electrically charged Reissner-Nordstrom-AdS solution in Einstein-Maxwell theory,
\eq
V(r)=\kappa-\frac{r_0^n}{r^n}+\frac{r^2}{\ell^2}+\frac{q_m^2}{2n(n+1)r^{2n}}
\label{magnetic p=n+1}\eeq
\eq
{\mc B}_{i_1\ldots i_{n+1}}=q_m\hat\ep_{i_1\ldots i_{n+1}},\qquad
H_{i_1\ldots i_{n+1}}=q_m\hat\ep_{i_1\ldots i_{n+1}}.
\eeq

\paragraph{Dyonic solution for $p=2$, $n=1$:} in four dimensions, the electric and magnetic fields of the previous two solutions can be combined in a single two-form field strength. This is again the familiar Reissner-Nordstrom-AdS solution in four dimensions, carrying both electric and magnetic charge,
\eq
V(r)=\kappa-\frac{r_0}{r}+\frac{r^2}{\ell^2}+\frac{q_e^2+q_m^2}{4r^2},
\qquad
H_{tr}=\frac{q_e}{r^{2}},
\qquad
H_{ij}=q_m\hat\ep_{ij}.
\label{dyonicRN}\eeq

\paragraph{Electric $p=n+3=D$ solution:} in this case $H$ is a spacetime filling field strength, and acts therefore as a cosmological constant. The solution is
\eq
V(r)=\kappa-\frac{r_0^n}{r^n}+\frac{r^2}{\ell^2}\lp1-\frac{q_e^2\ell^2}{2(n+1)(n+2)}\rp,
\label{electric p=n+3}\eeq
\eq
{\mc E}_{i_1\ldots i_{n+1}}=q_e\hat\ep_{i_1\ldots i_{n+1}},\qquad
H_{tri_1\ldots i_{n+1}}=q_er^{n+1}\hat\ep_{i_1\ldots i_{n+1}}.
\eeq
By tuning the electric charge $q_e$, it is possible to cancel completely the cosmological constant term from the lapse function, and obtain the Schwarzschild-Tangherlini solution \cite{Tangherlini:1963bw} when $\kappa=1$. Then the solution is asymptotically flat because the spacetime filling field strength also acts as a cosmological constant and cancels the effect of $\Lambda$ on the spacetime geometry.

\paragraph{Products of Einstein spaces with simple fluxes:} suppose now that $\mc H$ is the direct product of $N$ Einstein spaces $\mc H^{(a)}$, with induced metrics $\si^{(a)}_{ij}$ and Ricci tensors $\mc R^{(a)}_{ij}$, such that $R^{(a)}_{ij}=\kappa^{(a)}\si^{(a)}_{ij}$. Suppose that all $\kappa^{(a)}$ agree to a value that we will conventionally denote $n\kappa$, $n+1$ being the sum of the dimensions of the $\mc H^{(a)}$'s. Then the direct product $\mc H$ is also an Einstein space verifying  \eqref{Einstein manifold}. This opens up the possibility of having fluxes of $p$-form field strengths with a larger spectrum of ranks $p$.

Indeed, suppose the theory contains a $p$-form field $H_{[p]}$. As we have seen, it is defined by two polarization forms $\mc E$ and $\mc B$, of ranks $p-2$ and $p$ respectively. If all $\mc H^{(a)}$ factors have the same dimensionality $d$ (and hence $Nd=n+1$), we can turn on magnetic or electric fluxes of $H_{[p]}$ on every single Einstein space factor of $\mc H$ when $p=d$ or $p=d+2$ respectively, as follows.

Consider the magnetic case first. Suppose $\mc H$ is the direct product of $N$ $p$-dimensional Einstein manifolds; therefore $n=Np-1$. Each of these $\mc H^{(a)}$ supports its own volume form $\hat\ep^{(a)}_{[p]}$. Then,
\eq
\mc B_{[p]}=q_m\sum_{a=1}^N\pm\hat\ep^{(a)}
\label{Bprodmag}\eeq
solves\footnote{The equations constrain the magnitude of the flux to be equal on every single factor of the product of Einstein spaces, but leaves free the relative orientations, hence the arbitrary signs in \eqref{Bprodmag}.} equations \eqref{eqB} and \eqref{isotropy} as long as $p\geq2$ (we need at least two legs in the epsilons to solve \eqref{isotropy}, otherwise single terms in $\mc B_{[p]}$ mix and spoil the isotropy) and yields a genuine solution in $D=Np+2$ dimensions, with
\eq
\ep=\frac{Nq_m^2}{2r^{(2-N)p}},\qquad
P=\frac{2-N}2\frac{q_m^2}{r^{(2-N)p}},
\eeq
and hence
\eq
V(r)=\kappa-\frac{r_0^n}{r^n}+\frac{r^2}{\ell^2}-\frac{q_m^2}{2p(Np-2p+1)}\frac{1}{r^{2(p-1)}}.
\label{Vprodmag}\eeq
For $N=1$, we recover the previous magnetic $p=n+1$ solution \eqref{magnetic p=n+1}. 
Notice that, when $N\geq2$ the contribution of the $p$-form field to the lapse function changes sign and is always negative, meaning that there are regular black hole solutions even with $r_0=0$. However, in this case, the falloff of this term at large $r$ is slower than that of the mass term, and modifies the local asymptotic structure of the spacetime. Effectively, it behaves as a lower dimensional mass term. 

As a simple illustration of this construction, consider Einstein-Maxwell theory in $D=6$ dimensions. Then one has black hole solutions with $\mc S^2\times\mc S^2$ horizon, carrying magnetic flux through both spheres,
\eq
ds^2=-V(r)dt^2+\frac{dr^2}{V(r)}
+r^2\lp d\theta_1^2+\sin^2\theta_1\,d\phi_1^2\rp
+r^2\lp d\theta_2^2+\sin^2\theta_2\,d\phi_2^2\rp,
\label{S2xS2m}\eeq
\eq
F=q_m\lp\sin\theta_1\,\dd\theta_1\w\dd\phi_1\pm\sin\theta_2\,\dd\theta_2\w\dd\phi_2\rp,
\qquad V(r)=\frac13-\frac{r_0^3}{r^3}+\frac{r^2}{\ell^2}-\frac{q_m^2}{4r^2}.
\eeq
This magnetically charged solution first appeared in \cite{Maeda:2010qz}.  Let us switch off momentarily the cosmological constant for simplicity and without loss of generality. By a convenient rescaling of coordinates, $\rho=\sqrt{3} r$, we see that the $\mc S^2\times\mc S^2$ horizon gives rise to a five-dimensional `cone' over $\mc S^2\times\mc S^2$ with a solid angular deficit on the spheres{\footnote{It is interesting to note that, since $\mc S^2\times\mc S^2$ is not a homogeneous space, the geometry has a true curvature singularity at its apex $\rho=0$. However, going back to \eqref{S2xS2m}, this central singularity is hidden by a regular horizon due to the charge $q_m$, even if $r_0=0$.}}
\eq
dC^2=d\rho^2+\frac{\rho^2}{3}\lp d\theta_1^2+\sin^2\theta_1\,d\phi_1^2\rp
+ \frac{\rho^2}{3} \lp d\theta_2^2+\sin^2\theta_2\,d\phi_2^2\rp.
\eeq
Its uncharged Taub-NUT version  has been used to construct higher dimensional gravitational monopoles \cite{Mann}. 
This asymptotically conical space was shown to have a `balloon' type  instability \cite{kol} where one of the spheres inflates at the expense of the other. The solid angular deficit is provided by the $1/3$ value of the curvature term in the lapse function of (\ref{S2xS2m}). As a result each sphere has a reduced area of $4\pi r^2/3$  rather than $4\pi r^2$.  This is a typical property characterizing gravitational monopole solutions \cite{barriola} which we will encounter in all the solutions of this section.
A slightly more complicated example with a free three-form field strength living in eight dimensions, is given by a black hole solution with $\mc S^3\times\mc S^3$ horizon topology of the form
\eq
ds^2=-Vdt^2+\frac{dr^2}{V}
+r^2\lp d\theta_1^2+\sin^2\theta_1\,d\phi_1^2+\cos^2\theta_1\,d\psi_1^2\rp
+r^2\lp d\theta_2^2+\sin^2\theta_2\,d\phi_2^2+\cos^2\theta_2\,d\psi_2^2\rp,
\label{S3xS3m}\eeq
with
\eq
H=\frac{q_m}{2}\lp\sin2\theta_1\,\dd\theta_1\w\dd\phi_1\w\dd\psi_1
\pm\sin2\theta_2\,\dd\theta_2\w\dd\phi_2\w\dd\psi_2\rp,
\qquad V(r)=\frac25-\frac{r_0^5}{r^5}+\frac{r^2}{\ell^2}-\frac{q_m^2}{6r^4}.
\eeq
The area of each three-sphere is now reduced by the factor $2/5$ appearing in the lapse function of the black hole and the asymptotic space is therefore conical. In fact, it is interesting to note that for manifolds of the same topology, $\mc S^a\times\mc S^b$, with $a,b\geq 2$ and $5\leq a+b\leq 9$, infinitely many inhomogeneous metrics where shown to exist by B\"ohm \cite{bohm}. Black holes with B\"ohm type horizons were studied in \cite{gibbonshartnoll} where a balloon type instability was also encountered. The instability of horizons of this type is therefore quite generic and it is natural to question the fate of such an instability in the presence of magnetic charge. One could argue rather loosely that the magnetic charge may render rigid the horizon spheres.
We can understand intuitively the difference in-between the mass and charge terms in the black hole potential simply by observing that the three-forms source a stringlike object rather than a pointlike mass term, $r_0$. 
The space $\mc H$ can be exchanged with products of $\mathbb R^p$ spaces or $\mathbb H^p$ spaces with obvious modifications to $H$ and $V$.
Higher dimensional examples can be worked out trivially out of the general form \eqref{Bprodmag} and \eqref{Vprodmag} of the solution.

The electric case goes in the same way. Take $\mc H$ to be the direct product of $N$ $(p-2)$-dimensional Einstein spaces of same curvature, and put an equal electric flux of $H_{[p]}$ through each of the $\mc H^{(a)}$'s.
Again, to verify the isotropy condition \eqref{isotropy} we need $p\geq4$, and then we obtain
\eq
\mc E_{i_1\ldots i_{p-2}}=q_e\sum_{a=1}^N\pm\hat\ep^{(a)}_{i_1\ldots i_{p-2}},\qquad
\ep=\frac{Nq_e^2}{2r^{(N-2)p-2N+4}}
\eeq
\eq
V(r)=\kappa-\frac{r_0^n}{r^n}+\frac{r^2}{\ell^2}+\frac{q_e^2}{2(p-2)((N-2)(p-2)-1)}\frac1{r^{2(N-1)(p-2)-2}}.
\eeq
The charge contribution to $V$ is negative for $N<3$ and positive otherwise. When $N=1$, this is the electric $p=n+3$ solution of equation \eqref{electric p=n+3}. For $N=2$ these solutions are dual to the previous magnetic solutions \eqref{Vprodmag}, with a rank $p'=p-2$ field strength form, given in \eqref{Bprodmag}.
On the other hand, if one dualizes this solution with $N\geq3$, the resulting magnetic flux is not carried by a single Einstein space factor, but by a subset of them. This brings us to the next class of solutions.

\paragraph{Products of Einstein spaces with composite fluxes:}
if the flux of $\mc B$ or $\mc E$ is not carried by a single $\mc H^{(a)}$, but by two or more of them, the construction works the same, even if the dimension of $\mc H$ is not an integer multiple of the rank of the flux. Let $d\geq2$ be the dimension of the elementary spaces $\mc H^{(a)}$, and the rank $p$ of $\mc B$ an integer multiple of it, $p=md$. For any choice of $m$ elementary spaces $\mc H^{(a_1)}\times\cdots\times\mc H^{(a_m)}$, we have a volume $p$-form $\hat\ep^{\{a\}}=\hat\ep^{(a_1)}\w\ldots\w\hat\ep^{(a_m)}$ and we can define a flux
\eq
\mc B^{\{a\}}_{i_1\ldots i_p}=q_m\frac{p!}{(d!)^m}\hat\ep^{(a_1)}_{[i_1\ldots i_d}
\hat\ep^{(a_{2})}_{i_{d+1}\ldots i_{2d}}\cdots
\hat\ep^{(a_m)}_{i_{p-d+1}\ldots i_{p}]}.
\eeq
Here, with $\{a\}=\{a_1,\ldots,a_m\}$ we denote an ordered set of $m$ integers $1\leq a_1<\ldots< a_m\leq N$, corresponding to a choice of $m$ out of the $N$ elementary spaces $\mc H^{(a)}$. Such a flux breaks isotropy, because it picks $m$ privileged elementary spaces, but it can be restored by summing over all possible choices of the $m$ elementary spaces among the $N$ available, with same charge magnitude $|q_m|$,
\eq
\mc B_{i_1\ldots i_p}=\sum_{\{a\}}
\mc B^{\{a\}}_{i_1\ldots i_p}
=q_m\frac{p!}{(d!)^m}\sum_{\{a\}}\pm\hat\ep^{(a_1)}_{[i_1\ldots i_d}
\hat\ep^{(a_{2})}_{i_{d+1}\ldots i_{2d}}\cdots
\hat\ep^{(a_m)}_{i_{p-d+1}\ldots i_{p}]}.
\label{bbb}\eeq
This solves \eqref{isotropy}, and one therefore finds a magnetically charged solution. The orientation of the single fluxes remains arbitrary, but their strengths must match. A simple combinatorial calculation shows that with such a magnetic field we have
\eq
\mc B^2=\frac{p!N!}{m!(N-m)!}q_m^2.
\eeq
The geometry of this magnetically charged black hole is hence given by \eqref{metric}, with the lapse function (a logarithm appears when $2p=n+2$),
\eq
V(r)=\kappa-\frac{r_0^n}{r^n}+\frac{r^2}{\ell^2}+\frac{N!}{2(n+1)(2p-n-2)m!(N-m)!}\frac{q_m^2}{r^{2p-2}},
\label{Vbbb}\eeq
$\kappa$ being determined by the relation \eqref{Einstein manifold} on $\mc H$.

A simple example of this construction is in eight dimensions, with $\mc H=\mc S^2\times\mc S^2\times\mc S^2$ and a $p=4$ form field strength. Then, from \eqref{bbb}, we have the field strength
\eq
H=q_m\lp\hat\ep^{(1)}\w\hat\ep^{(2)}
\pm\hat\ep^{(2)}\w\hat\ep^{(3)}
\pm\hat\ep^{(1)}\w\hat\ep^{(3)}
\rp.
\eeq
Choosing coordinates such that
\eq
ds^2=-V(r)dt^2+\frac{dr^2}{V(r)}+r^2\lp
d\theta_1^2+\sin^2\theta_1\,d\phi_1^2
+d\theta_2^2+\sin^2\theta_2\,d\phi_2^2
+d\theta_3^2+\sin^2\theta_3\,d\phi_3^2
\rp,
\label{S3m}\eeq
we have a solution with
\eqa
&H=q_m\lp\sin\theta_1\sin\theta_2\,\dd\theta_1\w\dd\phi_1\w\dd\theta_2\w\dd\phi_2
\right.
&\qquad\quad V(r)=\frac{1}{5}-\frac{r_0^5}{r^5}+\frac{r^2}{\ell^2}+\frac{q_m^2}{4r^6},\qquad\qquad\nonumber\\
&\qquad\qquad
\pm\sin\theta_2\sin\theta_3\,\dd\theta_2\w\dd\phi_2\w\dd\theta_3\w\dd\phi_3\nonumber\\
&\qquad\qquad\qquad\left.
\pm\sin\theta_3\sin\theta_1\,\dd\theta_3\w\dd\phi_3\w\dd\theta_1\w\dd\phi_1
\rp.
\label{S3H}\eeqa

The extension of this construction for electric fluxes is straightforward.
Let $H_{[p]}$ be a $p$-form field of the form \eqref{H}, with purely electrical components, defined by a rank $p-2$ polarization form $\mc E$.
Now the dimension $d\geq2$ of the elementary spaces $\mc H^{(a)}$ must be an integer divisor of $p-2$, so that $p=md+2$. Again, any choice of $m$ elementary spaces $\mc H^{(a_1)}\times\cdots\times\mc H^{(a_m)}$ can carry the electric flux defined by
\eq
\mc E^{\{a\}}_{i_1\ldots i_{p-2}}=q_e\frac{(p-2)!}{(d!)^m}\hat\ep^{(a_1)}_{[i_1\ldots i_d}
\hat\ep^{(a_{2})}_{i_{d+1}\ldots i_{2d}}\cdots
\hat\ep^{(a_m)}_{i_{p-d-1}\ldots i_{p-2}]},
\eeq
with $\{a\}=\{a_1,\ldots,a_m\}$ defining the selection of elementary spaces $\mc H^{(a)}$ as before. Again, we restore isotropy by summing over all possible choices of the $m$ elementary spaces among the $N$ available, with same charge magnitude $|q_e|$,
\eq
\mc E_{i_1\ldots i_{p-2}}=\sum_{\{a\}}
\mc E^{\{a\}}_{i_1\ldots i_{p-2}}
=q_e\frac{(p-2)!}{(d!)^m}\sum_{\{a\}}\pm\hat\ep^{(a_1)}_{[i_1\ldots i_d}
\hat\ep^{(a_{2})}_{i_{d+1}\ldots i_{2d}}\cdots
\hat\ep^{(a_m)}_{i_{p-d-1}\ldots i_{p-2}]}.
\label{eee}\eeq
This solves \eqref{isotropy} with
\eq
\mc E^2=\frac{(p-2)!N!}{m!(N-m)!}q_e^2.
\eeq
and yields an electrically charged solution whose geometry is given by the metric \eqref{metric} and the lapse function (a logarithm appears when $2p=n+4$),
\eq
V(r)=\kappa-\frac{r_0^n}{r^n}+\frac{r^2}{\ell^2}+\frac{N!}{2(n+1)(n-2p+4)m!(N-m)!}\frac{q_e^2}{r^{2n-2p+4}},
\eeq
$\kappa$ being determined by the relation \eqref{Einstein manifold} on $\mc H$ as usual.
The dual of this solution, electrically charged under a $p=md+2$ form field strength, is a black hole that is magnetically charged under a dual rank $p'=D-p=m'd$ field strength, with $m'=N-m$ and charge $q_m=q_e$. This is precisely the solution given by equations \eqref{bbb} and \eqref{Vbbb}. In the particular case $m=N-1$, the dual reduces to the magnetic solution with simple fluxes given in equations \eqref{Bprodmag} and \eqref{Vprodmag}.

The previous eight-dimensional example with an Einstein space of the form $\mc H=\mc S^2\times\mc S^2\times\mc S^2$, can be extended to the electric case, with the same $p=4$ form field strength. The resulting $D=8$ black hole has metric \eqref{S3m} with
\eq
V(r)=\frac15-\frac{r_0^5}{r^5}+\frac{r^2}{\ell^2}+\frac{q_e^2}{4r^6},
\eeq
and a field strength
\eq
H=\frac{q_e}{r^2}\,\dd t\w\dd r\w\lp\sin\theta_1\,\dd\theta_1\w\dd\phi_1
+\sin\theta_2\,\dd\theta_2\w\dd\phi_2
+\sin\theta_3\,\dd\theta_3\w\dd\phi_3\rp.
\eeq
It so happens that in this case,  both electric and magnetic solutions, are built out of the same four-form field strength and this opens the door to dyonic solutions. We shall investigate these solutions in the next paragraph.

Another important observation is that for composite spaces of dimension $d=1$, the construction presented here fails. In Section~\ref{sec::multiple} we will show how to modify this construction by introducing more independent fields, in such a way that all field equations can be satisfied simultaneously when $\mc H$ is the direct product of $d=1$ dimensional spaces.

\paragraph{Dyonic black holes on products of two-dimensional Einstein spaces:}

In the construction above, we saw how to turn on electric or magnetic fluxes in $N$-fold products of Einstein manifolds of dimension $d$. In particular, the resulting black holes can be electrically charged if $p-2$ is an integer multiple of $d$, and magnetically charged if $p$ is an integer multiple of $d$. Hence when $d=2$, we can simultaneously turn on both magnetic and electric fluxes of an even $p$-form field if $N>p/2$, and the total field will simply be given by the sum of the electric and magnetic parts\footnote{The other possibility $d=1$ requires a flat $\mc H$ and is analyzed in the section~\ref{sec::multiple}.}. 
Then, the dyonic black hole solution is easily obtained by superposing the solutions of the previous paragraph and takes the form \eqref{metric} with (defining $m=p/2$),
\eq
V(r)=\kappa-\frac{r_0^n}{r^n}+\frac{r^2}{\ell^2}+
\frac{N!}{2(n+1)m!(N-m)!}\lp
\frac{1}{(2p-n-2)}\frac{q_m^2}{r^{2p-2}}+\frac{1}{(n-2p+4)}\frac{q_e^2}{r^{2n-2p+4}}
\rp,
\eeq
\eq
H=q_m\sum_{\{a\}}\pm\hat\ep^{(a_1)}\w\ldots\w\hat\ep^{(a_m)}
+\frac{q_e}{r^{n-2p+5}}\sum_{\{a\}}\pm\dd t\w\dd r\w\hat\ep^{(a_1)}\w\ldots\w\hat\ep^{(a_{m-1})}.
\eeq

One might wonder whether these dyonic solutions could enjoy (anti-)self-duality properties when $D=2p$. In this case, $N=p-1$ and the dimensionality of the spacetime must be a multiple of four, $D=4m$.
The previous solution simplifies to
\eq
V(r)=\kappa-\frac{r_0^n}{r^n}+\frac{r^2}{\ell^2}+\frac{(p-2)!}{m!(m-1)!}\frac{q_e^2+q_m^2}{4r^{n+1}},
\eeq
\eq
H=q_m\sum_{\{a\}}\pm\hat\ep^{(a_1)}\w\ldots\w\hat\ep^{(a_m)}
+\frac{q_e}{r^{2}}\sum_{\{a\}}\pm\dd t\w\dd r\w\hat\ep^{(a_1)}\w\ldots\w\hat\ep^{(a_{m-1})}.
\eeq
It can be readily checked that no choice of the relative signs can yield fields enjoying (anti-)self-duality properties in the Lorentzian signature. On the other hand, the associated Euclidean instanton, obtained by Wick rotating both time and electric charge is self-dual, provided the charges are equal in absolute value (i.e. taking $q_e=-iq_m$, with $q_m$ real) and all signs are taken to be equal. This condition enforces $q_e^2+q_m^2=0$ so that the function $V(r)$ coincides with the vacuum one \eqref{vacuum}. These instantons have the same vacuum AdS-bolt geometry as the Euclidean AdS black holes, but with non-trivial fields $H$. The latter have vanishing stress tensor and do not back react on the metric; they act as stealth fields.

Let us show how this works by writing down the simplest example, with $\mc H=\mc S^2\times\mc S^2\times\mc S^2$ and a $p=4$ form field strength. It is obtained by superposing the two eight-dimensional examples of the previous paragraph. The metric is given by \eqref{S3m}, with lapse function
\eq
V(r)=\frac15-\frac{r_0^5}{r^5}+\frac{r^2}{\ell^2}+\frac{q_e^2+q_m^2}{4r^6},
\eeq
and the total field strength reads
\eqa
&\displaystyle H=\frac{q_e}{r^2}\sin\theta_1\,\dd t\w\dd r\w\dd\theta_1\w\dd\phi_1+q_m\sin\theta_2\sin\theta_3\,\dd\theta_2\w\dd\phi_2\w\dd\theta_3\w\dd\phi_3
\nonumber\\
&\displaystyle\qquad\qquad\qquad
+\frac{q_e}{r^2}\sin\theta_2\,\dd t\w\dd r\w\dd\theta_2\w\dd\phi_2+q_m\sin\theta_3\sin\theta_1\,\dd\theta_3\w\dd\phi_3\w\dd\theta_1\w\dd\phi_1\nonumber\\
&\displaystyle\qquad\qquad\qquad\qquad\qquad
+\frac{q_e}{r^2}\sin\theta_3\,\dd t\w\dd r\w\dd\theta_3\w\dd\phi_3+q_m\sin\theta_1\sin\theta_2\,\dd\theta_1\w\dd\phi_1\w\dd\theta_2\w\dd\phi_2.
\eeqa
Now, analytically continue both the time and the electric charge to imaginary values, and impose $q_e=-iq_m$, to obtain an Euclidean solution with real components for the form field,
\eqa
&&ds^2=V(r)d\tau^2+\frac{dr^2}{V(r)}+r^2\lp
d\theta_1^2+\sin^2\theta_1\,d\phi_1^2
+d\theta_2^2+\sin^2\theta_2\,d\phi_2^2
+d\theta_3^2+\sin^2\theta_3\,d\phi_3^2
\rp,\nonumber\\
&&V(r)=\frac15-\frac{r_0^5}{r^5}+\frac{r^2}{\ell^2},
\label{Vsol}
\eeqa
and the total field strength reads
\eqa
&\displaystyle H=\frac{q_m}{r^2}\sin\theta_1\,\dd \tau\w\dd r\w\dd\theta_1\w\dd\phi_1+q_m\sin\theta_2\sin\theta_3\,\dd\theta_2\w\dd\phi_2\w\dd\theta_3\w\dd\phi_3
\nonumber\\
&\displaystyle\qquad\qquad\qquad
+\frac{q_m}{r^2}\sin\theta_2\,\dd\tau\w\dd r\w\dd\theta_2\w\dd\phi_2+q_m\sin\theta_3\sin\theta_1\,\dd\theta_3\w\dd\phi_3\w\dd\theta_1\w\dd\phi_1\nonumber\\
&\displaystyle\qquad\qquad\qquad\qquad\qquad
+\frac{q_m}{r^2}\sin\theta_3\,\dd\tau\w\dd r\w\dd\theta_3\w\dd\phi_3+q_m\sin\theta_1\sin\theta_2\,\dd\theta_1\w\dd\phi_1\w\dd\theta_2\w\dd\phi_2.
\eeqa
The topology of the Euclidean section is $\mathbb R^2\times\mc S^2\times\mc S^2\times\mc S^2$ with a bolt at the largest root of \eqref{Vsol}. Again each sphere $\mc S^2$ has reduced area given by $4\pi r^2/5$. The geometry is the same as the vacuum solution geometry, but there is an additional real self-dual $H$ field turned on, verifying $\star H =H$. In the next paragraph we shall look more closely to such Euclidean instantons.

\paragraph{Self-dual euclidean instantons for $2p=n+3$:} the reason why \eqref{dyonicisotropy} is not easily solved in Lorentzian signature when both the electric and magnetic parts of the field are switched on is the presence of an extra minus sign sitting in front of the terms quadratic in $\mc E$. It comes from the contractions in the time directions, and is usually an obstruction to Lorentzian dyonic solutions with a single field. However, if we choose to work with euclidean signature, starting with the metric
\eq
ds^2=V(r)d\tau^2+\frac{dr^2}{V(r)}+r^2\si_{ij}(y)dy^idy^j,
\label{emetric}\eeq
and the field $H_{[p]}$ given by \eqref{H}, the extra sign in the first equation in $\eqref{Hup}$ disappears, and the isotropy condition \eqref{dyonicisotropy} becomes
\eq
\mc B_{ik\ldots}\mc B_j{}^{k\ldots}-\frac{\mc B^2}{n+1}\si_{ij}=
-(p-1)(p-2)\lp\mc E_{ik\ldots}\mc E_j{}^{k\ldots}-\frac{\mc E^2}{n+1}\si_{ij}\rp.
\label{euciso}\eeq
We already found occurrences of such instantons in the previous paragraph, when the space $\mc H$ is the direct product of two dimensional spaces. Here we obtain new instantons with a different construction.

Take $\mc H$ to be the direct product of two Einstein manifolds $\mc H^{(1)}$ and $\mc H^{(2)}$ of dimensions $p-2$ and $p$ respectively, such that $\mc H$ is itself an Einstein manifold satisfying \eqref{Einstein manifold}. This is met if the metrics $\si^{(1)}_{ij}$ and $\si^{(2)}_{ij}$ and the Ricci tensors $\mc R^{(1)}_{ij}$ and $\mc R^{(2)}_{ij}$ of $\mc H^{(1)}$ and $\mc H^{(2)}$ are related by
\eq
\mc R^{(1)}_{ij}=n\kappa\si^{(1)}_{ij},\qquad
\mc R^{(2)}_{ij}=n\kappa\si^{(2)}_{ij}.
\eeq
Then, using the volumes form $\hat\ep^{(1)}$ and $\hat\ep^{(2)}$ on $\mc H^{(1)}$ and $\mc H^{(2)}$, we can generate on them an electric and a magnetic flux respectively, by taking,
\eq
\mc E=q_e\hat\ep^{(1)},\qquad
\mc B=q_m\hat\ep^{(2)}.
\eeq
By construction, these satisfy \eqref{eqE} and \eqref{eqB}, and it is easy to verify that if $q_e=q_m=q$, also \eqref{euciso} holds. A simple manipulation shows that for such a field the stress tensor vanishes everywhere, $T_{\mu\nu}=0$. 

The field $H_{[p]}$ does not back react on the metric, and acts as a stealth field. As a consequence, the geometry of these instantons coincides with the euclidean section of the vacuum AdS solutions with non-trivial topology: they have metric \eqref{metric} with lapse function \eqref{vacuum}. Note however that, when this solution is Wick rotated back to the Lorentzian signature, the electric charge becomes imaginary.

Finally, it is straightforward to check that the field strength of these solutions is self-dual, in the sense that,
\eq
H_{\mu_1\ldots\mu_p}=\frac{1}{p!}H_{\nu_1\ldots\nu_p}\hat\ep^{\nu_1\ldots\nu_p}{}_{\mu_1\ldots\mu_p},
\eeq
where $\hat\ep=r^{n+1}\dd\tau\w\dd r\w\hat\ep^{(1)}\w\hat\ep^{(2)}$ is the natural volume form of the euclidean space. When the rank $p$ is even (and therefore the dimension of the space a multiple of four) and $q_e=q_m$, this solution is self-dual, $\mc H=\star\mc H$.

As a simple example, we show the $D=8$ instanton with $p=4$ and transverse space $\mc H=\mc S^2\times\mc S^4$, with the relative radii of the spheres being chosen such that $\mc H$ is an Einstein space. The solution is self-dual and reads,
\eqa
&&ds^2=V(r)d\tau^2+\frac{dr^2}{V(r)}+r^2d\hat\Om^2_{(2)}+3r^2d\hat\Om^2_{(4)}\\
&&H=\frac q{r^2}\dd\tau\w\dd r\w\hat\ep_{[2]}+9q\,\hat\ep_{[4]},\qquad\qquad
V(r)=\frac15-\frac{r_0^5}{r^5}+\frac{r^2}{\ell^2},
\eeqa
where $d\hat\Omega^2_{(2)}$ and $d\hat\Omega^2_{(4)}$ are the line elements of the unit $\mc S^2$ and of the unit $\mc S^4$ respectively, and $\hat\ep_{[2]}$ and $\hat\ep_{[4]}$ the corresponding volume elements. The instanton has a solid deficit angle as that present in gravitational monopole solutions \cite{barriola},  without however a conical singularity since $r\geq r_h$ such that $V(r_h)=0$.

\section{Black holes with Einstein-K\"ahler horizons}\label{sec::kaehler}

If the spacetime is even dimensional, and the Einstein manifold $\mc H$ is also a K\"ahler space, there is an extra two form, the K\"ahler form $\om$. This form is harmonic, and so are the forms $\om^{(m)}$ of rank $2m$ obtained by taking the exterior product of $\om$ with itself $m$ times ($1\leq m\leq k$, with the dimension of $\mc H$ being $2k=n+1$). The two-form $\om$ also defines an almost complex structure on $\mc H$ and therefore meets by definition the isotropy condition \eqref{isotropy}. As a consequence, all $\om^{(m)}$ verify it, and we have a set of $k$ isotropic, harmonic forms on $\mc H$ that can be used to build solutions along the lines of what was done in the previous section with the volume form. Note that the maximum rank form $\om^{(k)}$ is proportional to the volume form on $\mc H$.

\paragraph{Magnetic Einstein-K\"ahler black hole:} consider a theory with an even $p$ form with $2\leq p\leq 2k$. Then the magnetic polarization form $\mc B=\om^{(p/2)}$, whose components are given by
\eq
\mc B_{i_1\ldots i_p}=q_m\frac{p!}{2^{p/2}}\om_{[i_1i_2}\ldots \om_{i_{p-1}i_p]},
\label{KB}\eeq
solves \eqref{eqB} and \eqref{isotropy}. Hence the geometry \eqref{metric} with lapse function obtained from equations \eqref{V} and \eqref{ep} and field strength \eqref{H} corresponding to the above $\mc B$, solves the equations of motion.

\paragraph{Electric Einstein-K\"ahler black hole:} the same construction works with electric fields. Consider again a theory with an even $p$-form, but with $4\leq p\leq 2k+2$. We can take as electric polarization form $\mc E=\om^{(p/2-1)}$, whose components are given by
\eq
\mc E_{i_1\ldots i_{p-2}}=q_e\frac{(p-2)!}{2^{p/2-1}}\om_{[i_1i_2}\ldots \om_{i_{p-3}i_{p-2}]},
\label{KE}\eeq
that solves both \eqref{eqB} and \eqref{isotropy}, and we obtain a solution with metric \eqref{metric}, \eqref{V} and \eqref{ep}, and field strength \eqref{H} corresponding to the above $\mc E$.

\paragraph{Dyonic Einstein-K\"ahler black holes:} by superposing the previous electric and magnetic solutions, we easily obtain dyonic solutions both electrically and magnetically charged under a $p$-form field strength such that $4\leq p\leq n+1$.

\paragraph{Direct products of Einstein-K\"ahler spaces:}
in the previous section, we explained how to construct more general solutions when $\mc H$ is a product of Einstein spaces. The same procedure can be carried out for products of Einstein-K\"ahler spaces, taking advantage of the various harmonic forms living on them. Observe that the  direct product of K\"ahler spaces is itself a K\"ahler space, whose K\"ahler form is given by the direct sum of the K\"ahler forms of the single factors. Therefore, for such an $\mc H$, solutions exist with the magnetic and electric polarizations given by \eqref{KB} and \eqref{KE} respectively. However, if the cohomology group of $\mc H$ allows it, more general fluxes can be built.

Indeed, having at hand isotropic and harmonic forms on the single factors of $\mc H$ allows to generalize the procedure we used to construct solutions with composite fluxes on products of Einstein spaces. All we have to do is to build the fields using the $\om^{(k)}$ forms on the single products instead of the volume forms. Very briefly, here is how it works.

Let $\mc H=\mc K^{(1)}\times\cdots\times\mc K^{(N)}$ be an Einstein space formed by the direct product of $N$ $d$-dimensional K\"ahler spaces $\mc K^{(a)}$ with associated K\"ahler forms $\hat\om^{(a)}$, and an even rank $p$-form field strength. Let $\hat\om^{(a)}$ be the K\"ahler form on $\mc K^{(a)}$. As before, we build the rank $2m$ harmonic forms $\hat\om^{(a,m)}$ on $\mc K^{(a)}$ as the $m$-fold exterior product of $\hat\om^{(a)}$ with itself.

If $2\leq p\leq d$, we can endow $\mc H$ with equal magnetic fluxes through each of the single K\"ahler factors,
\eq
\mc B_{[p]}=q_m\sum_{a=1}^N\pm\hat\om^{(a,p/2)}.
\label{KBprodmag}\eeq
When $4\leq p\leq d+2$ we obtain the electric counterpart of these solutions by taking
\eq
\mc E_{[p-2]}=q_e\sum_{a=1}^N\pm\hat\om^{(a,p/2-1)}.
\label{KEprodmag}\eeq
It is easy to show that these are isotropic harmonic forms on $\mc H$ as required, and the corresponding metric \eqref{metric}, \eqref{V} is obtained as usual. 

More general solutions can be obtained using `composite fluxes', as was done for Einstein space products, by decomposing the polarization forms into wedge products of harmonic forms and making the result isotropic by summing over all possible permutations. Rather than giving coumbersome general formulas, we shall illustrate it using a simple example in $D=10$, that can be easily reproduced in higher dimensions. Suppose $\mc H=\mc K^{(1)}\times\mc K^{(2)}$ is the direct product of two four dimensional Einstein-K\"ahler spaces, with K\"ahler forms $\hat\om^{(1)}$ and $\hat\om^{(2)}$ respectively. Then, we have the following possibilities for the fluxes,
\eq
\begin{array}{l@{\qquad}l}
p=2: &
\mc B =\hat\om^{(1)}\pm\hat\om^{(2)}\\
p=4: &
\mc B =\hat\om^{(1)}\w\,\hat\om^{(2)}\\
 &
\mc B =\hat\om^{(1)}\w\,\hat\om^{(1)}\pm\hat\om^{(2)}\w\,\hat\om^{(2)}\\
p=6: &
\mc B =\hat\om^{(1)}\w\,\hat\om^{(1)}\w\,\hat\om^{(2)}
\pm\hat\om^{(1)}\w\,\hat\om^{(2)}\w\,\hat\om^{(2)}\\
p=8: &
\mc B =\hat\om^{(1)}\w\,\hat\om^{(1)}\w\,\hat\om^{(2)}\w\,\hat\om^{(2)}
\end{array}
\eeq
Arbitrary linear combinations are possible in the $p=4$ case, the solution \eqref{KB} built out of the K\"ahler form of $\mc H$ being one,
\eq
\mc B=\lp\hat\om^{(1)}+\hat\om^{(2)}\rp\w\lp\hat\om^{(1)}+\hat\om^{(2)}\rp.
\eeq
The possible electric fluxes are obtained in a similar way.

A second example, with $D=14$ and $\mc H=\mc K^{(1)}\times\mc K^{(2)}\times\mc K^{(3)}$, the three factors being four dimensional Einstein-K\"ahler spaces, presents the following possibilities,
\eq
\begin{array}{l@{\qquad}l}
p=2: &
\mc B =\hat\om^{(1)}+\hat\om^{(2)}+\hat\om^{(3)}\\
p=4: &
\mc B =\hat\om^{(1)}\w\,\hat\om^{(2)}+\hat\om^{(2)}\w\,\hat\om^{(3)}+\hat\om^{(3)}\w\,\hat\om^{(1)}\\
&
\mc B =\hat\om^{(1)}\w\,\hat\om^{(1)}+\hat\om^{(2)}\w\,\hat\om^{(2)}+\hat\om^{(3)}\w\,\hat\om^{(3)}\\
p=6: &
\mc B =\hat\om^{(1)}\w\,\hat\om^{(2)}\w\,\hat\om^{(3)}\\
&
\mc B =\hat\om^{(1)}\w\,\hat\om^{(1)}\w\,\hat\om^{(2)}
+\hat\om^{(2)}\w\,\hat\om^{(2)}\w\,\hat\om^{(3)}
+\hat\om^{(3)}\w\,\hat\om^{(3)}\w\,\hat\om^{(1)}\\
p=8: &
\mc B =\hat\om^{(1)}\w\,\hat\om^{(1)}\w\,\hat\om^{(2)}\w\,\hat\om^{(3)}
+\hat\om^{(2)}\w\,\hat\om^{(2)}\w\,\hat\om^{(3)}\w\,\hat\om^{(1)}
+\hat\om^{(3)}\w\,\hat\om^{(3)}\w\,\hat\om^{(1)}\w\,\hat\om^{(2)}
\\
&
\mc B =\hat\om^{(1)}\w\,\hat\om^{(1)}\w\,\hat\om^{(2)}\w\,\hat\om^{(2)}
+\hat\om^{(2)}\w\,\hat\om^{(2)}\w\,\hat\om^{(3)}\w\,\hat\om^{(3)}
+\hat\om^{(3)}\w\,\hat\om^{(3)}\w\,\hat\om^{(1)}\w\,\hat\om^{(1)}\\
p=10: &
\mc B =\hat\om^{(1)}\w\,\hat\om^{(1)}\w\,\hat\om^{(2)}\w\,\hat\om^{(2)}\w\,\hat\om^{(3)}
+\hat\om^{(2)}\w\,\hat\om^{(2)}\w\,\hat\om^{(3)}\w\,\hat\om^{(3)}\w\,\hat\om^{(1)}\\
&\qquad\qquad
+\hat\om^{(3)}\w\,\hat\om^{(3)}\w\,\hat\om^{(1)}\w\,\hat\om^{(1)}\w\,\hat\om^{(2)}
\\
p=12:&
\mc B =\hat\om^{(1)}\w\,\hat\om^{(1)}\w\,\hat\om^{(2)}\w\,\hat\om^{(2)}
\w\,\hat\om^{(3)}\w\,\hat\om^{(3)}.
\end{array}
\eeq
The generalization to arbitrary products of Einstein-K\"ahler spaces is obvious.
The interested reader could build more explicit examples out of the Fubiny-Study metric on $\mathbb C{\mathrm P}^n$ or the Bergman metric on unit complex balls.

\section{Shaping black holes with multiple free fields}
\label{sec::multiple}

Consider a theory with two or more free $p$-form field strengths, possibly of different ranks. The total stress tensor is now the sum of the stress tensors of the single fields, and it can assume the form \eqref{st} even if the single fields break isotropy. Each field can be decomposed according to \eqref{H} into an electric and a magnetic part, that independently solve equations \eqref{eqE} and \eqref{eqB}. Then, once the isotropy condition is verified by the full stress tensor, the function $\ep(r)$ is simply the sum of the contributions of the single terms, and therefore $V(r)$ receives simple additive contributions. In the following, we will show how this can be easily achieved by accurately polarizing all fields.

The simplest way to enforce the isotropy condition is to have each field verifying independently equation \eqref{isotropy} or \eqref{dyonicisotropy}, in which case we can trivially superpose any two (or more) single field solutions.

\paragraph{Superposition principle: an example.}
Consider a theory with a two-form $F_{[2]}$ and a $(n+1)$-form $H_{[n+1]}$. Turning off the two-form we have the magnetic $p=n+1$ solution of the previous paragraph, while in absence of the $H$ field we obtain the electric $p=2$ solution. We can superpose these fields, and obtain black hole solutions electrically charged under $F$ and magnetically charged under $H$,
\eq
V=\kappa-\frac{r_0^n}{r^n}+\frac{r^2}{\ell^2}+\frac{q_e^2+q_m^2}{2n(n+1)r^{2n}},
\eeq
\eq
F_{tr}=\frac{q_e}{r^{n+1}},\qquad
H_{i_1\ldots i_{n+1}}=q_m\hat\ep_{i_1\ldots i_{n+1}}.
\eeq
In particular, since the electric components of a two-form have no legs in $\mc H$, such an electric charge can be added to all the solutions that we present in this article, although we will not always display it explicitly.\\

More generally, when the rank $p$ of a form field does not match the dimension of $\mc H$, it will automatically introduce some privileged directions in $\mc H$, and its stress tensor will not be of the form \eqref{st}. However, it is possible to take multiple copies of the field and orient them in such a way that the full stress tensor of the combined fields is in the correct form, as we show next. While this construction might in principle work with curved Einstein spaces, it requires to find all harmonic forms on them, a challenging enterprise. We will limit the analysis of such solutions in the following sections to flat $\mc H$.

\paragraph{Isotropy from multiple fields on $\mathbb R^{n+1}$: the magnetic case.}
Suppose $1\leq p\leq n+1$, and let $E^a{}_i$ be an orthonormal basis of dual vector fields on $\mathbb R^{n+1}$. Here the index $a=1,\ldots n+1$ labels these  vectors, that collectively form a vielbein $E^a=E^a{}_i\,\dd y^i$. 
By definition, we have the relations $\si^{ij}E^a{}_iE^b{}_j=\delta^{ab}$ and $\si_{ij}=\delta_{ab}E^a{}_iE^b{}_j$.
Consider now the magnetic part of a $p$ form field strength $H_{[p]}$. It has $p$ legs, and it is not possible to turn on some flux on $\mc H$ without breaking isotropy. However, one can build a stress tensor of the form \eqref{st} out of 
\eq
N_b=\frac{(n+1)!}{p!(n-p+1)!}
\eeq
independent free $p$-form fields, whose legs are distributed in such a way that the total stress tensor, being the sum of the stress tensors of the single fields, recovers isotropy and is of the form \eqref{st}.
The construction goes as follows.
We can label the single field with an ordered set $\{a\}=\{a_1,\ldots,a_{p}\}$ of integers such that $1\leq a_1<\ldots<a_{p}\leq n+1$. These integers define the directions in which the legs of the corresponding field lie, according to\footnote{In the language of differential forms, $\mc B^{\{a\}}=q_m E^{a_1} \wedge \ldots \wedge E^{a_p}$.} 
\eq
\mc B^{\{a_1,\ldots,a_{p}\}}_{i_1\ldots i_{p}}=q_mp!E^{a_1}{}_{[i_1}\cdots
E^{a_p}{}_{i_p]}.
\label{ban}\eeq
This tensor satisfies trivially \eqref{eqB} with a flat induced metric on $\mc H$. Then, the contraction
\eq
\mc B^{\{a\}}_{ik_2\ldots k_p}\mc B^{\{a\}}_{j}{}^{k_2\ldots k_p}=
q_m^2(p-1)!\sum_{l=1}^p E^{a_l}{}_iE^{a_l}{}_j
\eeq
becomes isotropic once the sum over all field labels $\{a\}$ is performed,
\eq
\sum_{\{a\}}\mc B^{\{a\}}_{ik_2\ldots k_p}\mc B^{\{a\}}_{j}{}^{k_2\ldots k_p}=
q_m^2\frac{n!}{(n-p+1)!}\si_{ij}.
\eeq
It follows then that the stress tensor assumes the isotropic form \eqref{st} with
\eq
\ep(r)=\frac{(n+1)!}{2p!(n-p+1)!}\frac{q_m^2}{r^{2p-n-1}},\qquad
P(r)=\frac{n!(2p-n-1)}{2p!(n-p+1)!}\frac{q_m^2}{r^{2p-n-1}},
\eeq
and the lapse function becomes (a logarithm appears when $2p= n+2$),
\eq
V(r)=-\frac{r_0^n}{r^n}+\frac{r^2}{\ell^2}+\frac{n!}{2p!(n-p+1)!(2p-n-2)}
\frac{q_m^2}{r^{2p-2}},
\eeq
since for a flat $\mc H$ we have $\kappa=0$.
When $p=n+1$ we have one single magnetic field and we recover the result \eqref{magnetic p=n+1} in the particular $\kappa=0$ case, for which $\mc H=\mathbb R^{n+1}$.

A particularly interesting case comes about for $p=1$. The corresponding solutions display $n+1$ scalar fields $\phi^{(i)}$, one for each coordinate of the flat transverse space, and the $r$-dependence drops from the extra contribution in the lapse function due to the charges. As a result we have the solution,
\eq
ds^2=-V(r)dt^2+\frac{dr^2}{V(r)}+r^2\sum_{i=1}^{n+1}\lp dy^i\rp^2,\qquad
V(r)=-\frac{q_m^2}{2n}-\frac{r_0^n}{r^n}+\frac{r^2}{\ell^2},\qquad
\phi^{(i)}=q_my^i,
\label{magscalar}\eeq
describing an AdS black hole with flat horizon, but with a lapse function of the form usually associated to black holes with an hyperbolic horizon \cite{Birmingham:1998nr}, with an effective $\kappa_{\rm eff}=-q_m^2/2n<0$. Now, noticing that the scalar fields enter the action only through their derivatives and therefore enjoy a shift symmetry, one can argue that these scalar fields are only defined up to a constant. This is the scalar analogue of the gauge invariance of the $p$-form fields. In this case, one can compactify the horizon to a $(n+1)$-dimensional torus, because the discontinuities of the scalars on the identifications can be gauged away using their shift symmetry. This yields asymptotically locally AdS black holes with a toroidal horizon dressed by scalar fields, and -- in four dimensions, where these scalars are axions -- it is precisely the axionic black hole that we will analyze in the the next section. Finally, we will come back on the question of the shift symmetry of the scalars in the concluding section of the article.

\paragraph{Isotropy from multiple fields on $\mathbb R^{n+1}$: the electric case.} The same construction can be carried out in the electric case, yielding the dual solution to the previous one. Suppose $3\leq p\leq n+3$, and the matter sector of the theory has
\eq
N_e=\frac{(n+1)!}{(p-2)!(n-p+3)!}
\eeq
$p$-form field strengths. Labeling the fields with an ordered set $\{a\}=\{a_1,\ldots,a_{p-2}\}$ of integers such that $1\leq a_1<\ldots<a_{p-2}\leq n+1$ as before, we can define the electric polarization vectors
\eq
\mc E^{\{a_1,\ldots,a_{p-2}\}}_{i_1\ldots i_{p-2}}=q_e(p-2)!E^{a_1}{}_{[i_1}\cdots
E^{a_{p-2}}{}_{i_{p-2}]}.
\label{ean}\eeq
Then, the contraction
\eq
\mc E^{\{a\}}_{ik_2\ldots k_{p-2}}\mc E^{\{a\}}_{j}{}^{k_2\ldots k_{p-2}}=
q_e^2(p-3)!\sum_{l=1}^{p-2} E^{a_l}{}_iE^{a_l}{}_j
\eeq
becomes isotropic once the sum over all field labels $\{a\}$ is performed,
\eq
\sum_{\{a\}}\mc E^{\{a\}}_{ik_2\ldots k_p}\mc E^{\{a\}}_{j}{}^{k_2\ldots k_p}=
q_e^2\frac{n!}{(n-p+3)!}\si_{ij}.
\eeq
It follows then that the stress tensor assumes the isotropic form \eqref{st} with
\eq
\ep(r)=\frac{(n+1)!}{2(p-2)!(n-p+3)!}\frac{q_e^2}{r^{n-2p+5}},\qquad
P(r)=\frac{n!(n-2p+5)}{2(p-2)!(n-p+3)!}\frac{q_e^2}{r^{n-2p+5}},
\eeq
and the lapse function becomes (a logarithm appears when $2p=n+4$),
\eq
V(r)=-\frac{r_0^n}{r^n}+\frac{r^2}{\ell^2}+\frac{n!}{2(p-2)!(n-p+3)!(n-2p+4)}
\frac{q_e^2}{r^{2n-2p+4}},
\eeq
since for a flat $\mc H$ we have $\kappa=0$.
When $p=n+3$ we have one single electric field and we recover the result \eqref{electric p=n+3} in the particular $\kappa=0$ case, for which $\mc H=\mathbb R^{n+1}$. Finally, for $p=n+2$ we retrieve the dual solution to the black hole with $n+1$ `magnetic' scalar fields \eqref{magscalar} of the previous paragraph, with the metric determined by the function
\eq
\label{axion1}
V(r)=-\frac{q_e^2}{2n}-\frac{r_0^n}{r^n}+\frac{r^2}{\ell^2},
\eeq
Again, the resulting spacetime defines an AdS black hole with flat horizon, but with a lapse function of the form usually associated to hyperbolic black holes, with an effective $\kappa_{\rm eff}=-q_e^2/2n<0$.

More generally, the electromagnetic duality links the solution with electric $p$-forms $H_{[p]}^{\{a\}}$ with polarizations \eqref{ean} to the solution obtained in the previous paragraph, with magnetic $p_\star$-forms $H_{[p_\star]}=\star H_{[p]}$ polarized according to \eqref{ban} with $q_e=q_m$, where $p_\star=n+3-p$. It can be easily checked that the number of fields match, i.e. $N_e(n,p)=N_m(n,p_\star)$.

\paragraph{Dyonic solutions on $\mathbb R^{n+1}$ with $3\leq p\leq n+1$.}
For this range of $p$, we have obtained both a magnetic solution with $N_b$ fields and an electric solution with $N_e$ fields. Using the superposition principle we can have both electric and magnetic fluxes. Consider for example $N_b+N_e$ $p$-form fields, the first $N_b$ of the form \eqref{ban} and the rest of the form \eqref{ean}. Then we have a solution with
\eq
V(r)=-\frac{r_0^n}{r^n}+\frac{r^2}{\ell^2}+\frac{n!}{2p!(n-p+1)!(2p-n-2)}
\frac{q_m^2}{r^{2p-2}}
+\frac{n!}{2(p-2)!(n-p+3)!(n-2p+4)}
\frac{q_e^2}{r^{2n-2p+4}}.
\eeq
A particularly interesting case arises when $2p=n+3$. In this case the number $N$ of electric and magnetic fluxes we need coincide,
\eq
N=N_b=N_e=\frac{(2p-2)!}{p!(p-2)!},
\eeq
and count the number of ways one can pick $p-2$ indices out of $2p-2=n+1$. We can therefore consider $N$ fields with both electric and magnetic parts turned on. We take one field for every choice of $p-2$ indices out of $n+1$, and we put the electric legs along those directions, and the magnetic ones in the remaining $p$. The explicit form of the field polarizations, if we label them again with ordered sets $\{a\}$ of $p-2$ integers, is given by
\eq
\mc E^{\{a_1,\ldots,a_{p-2}\}}_{i_1\ldots i_{p-2}}=q_e(p-2)!E^{a_1}{}_{[i_1}\cdots
E^{a_{p-2}}{}_{i_{p-2}]},\quad
\mc B^{\{a_1,\ldots,a_{p-2}\}}_{i_1\ldots i_{p}}=q_m\varepsilon^{a_1\ldots a_{p-2}b_1\ldots b_{p}}E^{b_1}{}_{[i_1}\cdots
E^{b_p}{}_{i_p]}. 
\eeq
where $\varepsilon^{a_1\ldots a_{n+1}}$ is the totally antisymmetric tensor with components $0$, $\pm1$ according to the sign of the permutation of the indices, and a sum over the repeated $b_i$ indices is understood. The resulting spacetime has
\eq
V(r)=-\frac{r_0^n}{r^n}+\frac{r^2}{\ell^2}+\frac{N}{2(n+1)}\frac{q_e^2+q_m^2}{r^{n+1}},
\eeq
with the field components given by
\eq
H^{\{a\}}_{tri_1\ldots i_{p-2}}=\frac1{r^2}\mc E^{\{a\}}_{i_1\ldots i_{p-2}},\qquad
H^{\{a\}}_{i_1\ldots i_{p}}=\mc B^{\{a\}}_{i_1\ldots i_{p}}.
\eeq
When $p$ is odd, this solution is anti-self-dual when $q_e=q_m$, and self-dual for $q_e=-q_m$.

Note that these black holes generalize the dyonic Reissner-Nordstrom-AdS solution in four dimensions. The latter, given in \eqref{dyonicRN}, is indeed recovered by setting $n=1$ and $p=2$ in the previous expressions.

In terms of forms they assume a particularly simple expression. Define $E^a=E^a{}_idy^i$. Then
\eq
H^{\{a\}}=\frac{q_e}{r^2}\,\dd t\w\dd r\w E^{a_1}\w\ldots\w E^{a_{p-2}}
+\frac{1}{p!}q_m\varepsilon^{a_1\ldots a_{p-2}b_1\ldots b_p}
E^{b_1}\w\ldots\w E^{b_p}.
\eeq
Moreover, choosing cartesian coordinates on $\mc H$, such that $\si_{ij}=\delta_{ij}$, we can choose a gauge in which $E^{a}=\dd y^a$. For illustrative purposes, we present the six dimensional case, for which $n=p=3$. We need in this case $N=4$ three-form fields. The dyonic solution is then given by
\eqa
&\displaystyle ds^2=-V(r)dt^2+\frac{dr^2}{V(r)}+r^2\sum_{i=1}^{4}(dy^i)^2,\qquad
&V(r)=-\frac{r_0^3}{r^3}+\frac{r^2}{\ell^2}+\frac{q_e^2+q_m^2}{2r^{4}},\\
&\displaystyle H^{\{1\}}=\frac{q_e}{r^2}\,\dd t\w\dd r\w\dd y^1+q_m\,\dd y^2\w\dd y^3\w\dd y^4,\qquad
&H^{\{2\}}=\frac{q_e}{r^2}\,\dd t\w\dd r\w\dd y^2+q_m\,\dd y^3\w\dd y^4\w\dd y^1,\nonumber\\
&\displaystyle H^{\{3\}}=\frac{q_e}{r^2}\,\dd t\w\dd r\w\dd y^3+q_m\,\dd y^4\w\dd y^1\w\dd y^2,\qquad
&H^{\{4\}}=\frac{q_e}{r^2}\,\dd t\w\dd r\w\dd y^4+q_m\,\dd y^1\w\dd y^2\w\dd y^3,\nonumber
\eeqa
and is self-dual when $q_e=-q_m$, anti-selfdual when $q_e=q_m$.

This spacetime contains a black hole with an electric and magnetic three-form charge dressing. The solution has similar structure to the planar six dimensional Reissner-Nordstrom black hole, although here the three-form has a slower fall off in the radial coordinate $r$, ($r^{-4}$ rather than an $r^{-6}$). The inverse temperature is given by,
\beq
\beta=\frac{4 \pi}{|V'(r_h)|}=\frac{8 \pi\ell^2 r_h^5}{10 r_h^6-(q_e^2+q_m^2)\ell^2}
\eeq
The solution is a regular black hole with an inner Cauchy horizon and an outer event horizon as long as, $r_0^3\leq \frac{3(q_e^2+q_m^2)}{5r_h}$. This bound is saturated for the extremal black hole with $r_{h,ext}^3=(q_e^2+q_m^2)\ell^2/10$.\\

Obviously, this procedure can be generalized to superpose multiple forms of ranks ranging from $3$ to $n+1$. This allows to shape the lapse function with an even power series ranging from $r^2$ to $1/r^{2n}$,
\eq
V(r)=-\frac{r_0^n}{r^n}+\frac{r^2}{\ell^2}+\sum_{m=1}^n\frac{c_m}{r^{2m}},
\eeq
the $c_m$ being constants determined by the charges. Additional $\log r$ terms can appear for forms verifying $2p=n+2$ or $2p=n+4$.


\section{Axionic black holes and their extensions}\label{sec::axion}

In this section we will briefly revisit the case of black holes with a three form dressing encountered in the previous section. We shall focus  on the four dimensional case since this corresponds precisely to an axionic black hole. This, to our knowledge, is the first static black hole in the literature presenting non-trivial axionic charge. Solutions of Einstein's equations in presence of these fields were explored for the first time in \cite{Bowick:1988xh} and \cite{Xanthopoulos:1991mx} but in the first case the charge was zero whereas in the second case the solutions were singular.

Consider the toroidal black hole,
\beq
ds^2=-V(r)dt^2+\frac{dr^2}{V(r)}+r^2(dx^2+dy^2),
\label{tormet}\eeq
dressed with two constant electric three-forms,
\beq
H^{(1)}=q_e \dd t \wedge \dd r \wedge \dd x, \qquad H^{(2)}=q_e \dd t \wedge \dd r \wedge \dd y,
\eeq
and resulting lapse function,
\beq
V(r)=-\frac{q_e^2}{2} + \frac{r^2}{\ell^2}-\frac{r_0}{r}.
\label{axioniclapse}\eeq
As noted above (\ref{axion1}) the constant axionic charge $q_e$ is associated to a would be horizon curvature term (the charge can be magnetic and related to (axionic) scalars (\ref{magscalar})). Just like for black holes with hyperbolic horizons, we have to consider a negative cosmological constant in order to avoid a naked singularity at $r=0$. Therefore, we have at hand an asymptotically locally AdS black hole with a planar horizon whose axionic charges render its properties similar to those of the uncharged hyperbolic 
black hole (see for example \cite{Emparan:1999gf}).

For a start the inverse temperature reads,
\beq
\label{conical}
\beta=\frac{4 \pi}{|V'(r_h)|}=\frac{4 \pi\ell^2 r_h}{3 r_h^2-\half q_e^2\ell^2},
\eeq
where we have replaced the mass parameter $r_0=r_h\lp\frac{r_h^2}{\ell^2}-\half q_e^2\rp$ by the outermost root of the lapse function, the event horizon at $r=r_h$.
Recall that in a Reissner-Nordstrom spacetime (planar or spherical) switching off the mass parameter yields a singular spacetime. In fact, beyond the extremal bound Reissner-Nordstrom spacetimes are always singular. The effect of the axionic charges is quite different. For a start setting $r_0=0$ the axion solution is still a regular black hole with a horizon at $r_h=\frac{q_e l}{\sqrt{2}}$ and a curvature singularity at $r=0$. The event horizon is  supported by the axionic charges providing the necessary  scale to form the horizon. Again, just like for hyperbolic black holes, for negative mass, $r_0<0$, we have an inner Cauchy horizon with an extremal black hole attained for mass parameter,
\beq
r_0^{ext}=-\frac{|q_e|^3\ell}{3 \sqrt{6}},
\eeq
and unique event horizon at $r_{ext}=\frac{|q_e|\ell}{\sqrt{6}}$. In other words for $r_0^{ext}<r_0<0$ we have a regular black hole with an inner Cauchy horizon and an outer event horizon. Hence the axion fields permit not only smaller $r_0$ black holes, but also negative mass planar black holes, a possibility that is usually associated to hyperbolic horizons in vacuum AdS. In fact, the bigger the axionic charge, the bigger in magnitude the negative mass black hole that spacetime can support. Actually, we can go a bit further and now add a Maxwell field carrying an electromagnetic charge, say $Q$, to our axionic black hole,
\beq
V(r)=-\frac{q_e^2}{2} + \frac{r^2}{\ell^2}-\frac{r_0}{r}+\frac{Q^2}{r^2}.
\eeq
We see immediately that adding axionic charges to a planar  Reissner-Nordstrom black hole can lead to permissible smaller mass black holes. To check this, we can even switch off the mass parameter $r_0$. The lapse function then reads,
\beq
V(r)=-\frac{q_e^2}{2} + \frac{r^2}{\ell^2}+\frac{Q^2}{r^2},
\eeq
and this spacetime can have two horizons, an inner Cauchy horizon and an outer event horizon at
\beq
r_h^2=\frac{q_e^2\ell^2\pm \sqrt{q_e^4\ell^4-16Q^2\ell^2}}{4},
\eeq
as long as the axionic charge is such that $q_e^2\geq \frac{4|Q|}{\ell}$. There is an extremal black hole for $r_h=\frac{|q_e|\ell}{2}$. Adding mass does not change these results qualitatively since the mass term $r_0$ also acts as a regulating term. The same properties hold in higher dimension as long as $p=n+2$, hence four-forms black holes  in five dimensional spacetime etc. In a nutshell, axionic charges operate as a negative curvature term and have the tendency to regularize the geometry of spacetime. 

Another interesting solution is obtained by Wick rotation of the above metric (\ref{axioniclapse}). Let us consider the following imaginary transformation $t\rightarrow i \theta$, $x\rightarrow -i \tau$, $q_e\rightarrow i q_s$. The transformed metric has again real components and reads,
\beq
\label{straight}
ds^2=V(r)d\theta^2+\frac{dr^2}{V(r)}+r^2(-d\tau^2+dy^2)
\eeq
with lapse function $V(r)=\frac{q_s^2}{2}+\frac{r^2}{\ell^2}-\frac{r_0}{r}$. The $r$ coordinate has range, $r_h\leq r$ and the metric has now an axial Killing vector $\partial_\theta$ with the axis at $r=r_h$. The azimuthal angle $\theta$ has a deficit angle $\Delta=2\pi (1-\beta)$ provided via the conical singularity{\footnote{By a convenient identification of the $\theta$ coordinate we can always do away with one conical singularity but here we keep it in order to account for the cosmic string.}} at $r=r_h$ (\ref{conical}). This wedge is accounted for by the presence of an infinitesimally thin cosmic string whose worldsheet lies on the $r=r_h$ two-plane defined by $V(r_h)=0$ (see for example \cite{vilenkin}). The core of the string is sourced by a distributional energy-momentum tensor $T_\mu{}^\nu=-\delta^{(2)} T \delta_\mu{}^\nu$ of string tension $T=\frac{\Delta}{8\pi G}$.
The axionic scalars (dual to the two three-form fields) combine into a time dependent single complex field, $\Phi=q_s(\tau+i y)$ with charge $q_s$ running along the string direction, in this way we get a solution of the following action
\beq S=\frac{1}{16\pi G} \int d^4 x \sqrt{-g} \left( R - 2\Lambda - \frac{1}{2}\partial_\mu\Phi \partial^\mu\Phi^* \right) \ . \eeq
Note here that the axionic charge has changed overall sign in the lapse function and as a result we can have flat, de~Sitter or anti-de~Sitter asymptotic geometries. Focusing on the $\Lambda=0$ case we find that the string tension is,
\beq
T=\frac{q_s^4-16\pi r_0}{4 G q_s^4}
\eeq
Global, unlike local strings, have long range interactions due to the presence, at low energy scales, of an axionic field related to the remnant Goldstone boson \cite{vilenkin}. The gravitational field of straight global strings, as the one pictured in (\ref{straight}), has been argued to be singular \cite{cohenkaplan}. However, if one allows for the strings to intrinsically inflate -- quite like domain wall spacetimes \cite{sikivie} -- Gregory \cite{gregory} elegantly argued that the singularity may be swept away. One is tempted to interpret this metric in relation to the far off gravitational field of a global string but this to our understanding is a quite separate and non trivial question going beyond the scope of this paper. 

Finally, we would like to point out another simple generalization of the black holes presented in this article. Switching to Eddington-Finkelstein coordinates, the axionic black hole reads
\beqa
&&ds^2=-V(r)du^2-2dr du+r^2\lp dx^2+dy^2\rp,\\
&&H^{(1)}=q_e \dd u \wedge \dd r \wedge\dd x,\qquad
H^{(2)}=q_e \dd u \wedge \dd r \wedge\dd y,\nonumber
\eeqa
with the lapse function $V(r)$ still given by \eqref{axioniclapse}.
Dropping the stationarity hypothesis and adding some external matter sourcing a null stress tensor $T^{\mathrm{ext}}_{uu}=\mu(u,r)/8\pi G$, and enhancing $r_0=r_0(u)$ to an arbitrary function of the retarded (or anticipated) coordinate $u$, one obtains a Vaidya type solution \cite{Vaidya:1999zz} representing a radiating axionic black hole as long as
\eq
\mu(u,r)=-\frac{\p_ur_0}{r^2}
\eeq
is verified. In four dimensions, the null stress tensor $T^{\mathrm{ext}}_{uu}$ can be generated by a Maxwell field $F_{[2]}$, but such radiating solutions can be easily build for any other black hole presented here.

\section{Thermodynamical properties and phases of the black holes}
\label{sec::thermo}

In the previous sections, we constructed a large class of black holes dressed by one or more $p$-form fields. They are solution of the following action,
\beq S = S_0 + \sum_{p=1}^D \sum_{k=1}^{N_p} S_M^{(p,k)},
\label{action1}\eeq
where $S_0$ is the Einstein-Hilbert action with cosmological constant given in \eqref{action}, $N_p$ is the number of $p$-form fields $H_{[p]}^{(k)}$, and the matter sector consists in the sum of the single free field contributions of the form \eqref{SM},
\beq
 S_M^{(p,k)} = -\frac{1}{16\pi G} \int_{\mc M}  d^Dx \sqrt{-g} \frac{1}{2p!} \lp H_{[p]}^{(k)}\rp^2. 
\eeq
We have shown that static solutions of such theories have a metric of the form \eqref{metric} with lapse function \eqref{V} determined by
\beq 
\ep(r)=\sum_{p,k} \ep_{(p,k)}(r),\qquad
\ep_{(p,k)}(r) =  \frac{\mc E^{(k)2}_{[p-2]}}{2(p-2)! r^{n-2p+5}} +  \frac{\mc B^{(k)2}_{[p]}}{2p! r^{2p-n-1}},
\eeq
as a consequence of the superposition principle previously discussed. Hence the full expression for the black hole potential is,
\beq 
V(r) = \kappa - \frac{r_0^n}{r^n} + \frac{r^2}{\ell^2} + \frac{1}{2(n+1)} \sum_{p,k} \left[ \frac{\mc E^{(k)2}_{[p-2]}}{(p-2)! (n-2p+4) r^{2(n-p+2)}} +  \frac{\mc B^{(k)2}_{[p]}}{p! (2p-n-2) r^{2(p-1)}}  \right].
\label{genV}\eeq
In order to simplify the discussion, we have omitted the logarithmic terms appearing when $2p=n+4$ and $2p=n+2$ in an odd dimensional spacetime. The reader can easily reintroduce them when needed. The shape of the electric part of this series ranges from $1/r^{2n}$ when $p=2$ to $r^2$ when $p=n+3$, like the cosmological constant term. As for the magnetic part it ranges from $1/r^0$ when $p=1$, like the curvature term, to $1/r^{2n}$ when $p=n+1$.

If $r_0$ is large enough, the function $\eqref{genV}$ will always have at least one positive root. Let us call $r_h$ the largest of these roots. The solution will then exhibit an event horizon located at $r=r_h$, and the spacetime will contain a black hole. Its temperature, proportional to surface gravity of its outermost horizon, is given by
\beq 
T=\frac{\left|V'(r_h)\right|}{4\pi} =  \frac{n\kappa}{4 \pi r_h} + \frac{(n+2)r_h}{4\pi\ell^2} - \frac{1}{8\pi (n+1)} \sum_{p,k} \left( \frac{\mc E^{(k)2}_{[p-2]}}{(p-2)! r_h^{2n-2p+5}} +  \frac{\mc B^{(k)2}_{[p]}}{p! r_h^{2p-1}} \right).
\eeq

For reasons of clarity, we will now focus our attention on the electric case and we turn on all rank $p$ fields strengths that satisfy
$2p\leq n+3$. When this inequality holds, the corresponding field strength vanishes at spatial infinity and allows to obtain finite thermodynamical potentials after background subtraction, as we shall shortly see. As a result, we associate a rank $p-1$ electric potential form to each  $p$-form field by $H_{[p]}^{(k)}=\dd \mc A_{[p-1]}^{(k)}$.  A possible choice is
\beq
\mc A^{(k)}_{t i_1\ldots i_{p-2}} = \Phi^{(k)}_{i_1 \ldots i_{p-2}} + \frac{1}{(n-2p+4)r^{n-2p+4}} \mc E^{(k)}_{i_1 \ldots i_{p-2}}\,.
\eeq
Here $ \Phi^{(k)}_{i_1 \ldots i_{p-2}}$ provides the value of the electric potential at spatial infinity. Then, we fix the gauge by imposing that the  electric potentials vanish on the outermost horizon $r_h$ by taking
\beq
\Phi^{(k)}_{i_1 \ldots i_{p-2}} = - \frac{1}{(n-2p+4)r_h^{n-2p+4}} \mc E^{(k)}_{i_1 \ldots i_{p-2}} \ ,
\eeq
Equivalently, one can ask for the regularity of the scalar quantity $\lp\mc A_{[p-1]}^{(k)}\rp^2$ on the horizon, as first discussed in \cite{Gibbons:1976ue} for the electromagnetic case ($p=2$). 
In our restricted case we find,
\beq
T = \frac{n\kappa}{4 \pi r_h} + \frac{(n+2)r_h}{4\pi l^2} - \frac{1}{8\pi (n+1)} \sum_{p,k}  (n-2p+4)^2 r_h^{3-2p} \frac{\lp\Phi^{(k)}_{[p-2]}\rp^2}{(p-2)!}.
\label{genT}\eeq
In the grand canonical ensemble, $r_h=r_h(M,\Phi^{(k)}_{[p-2]})$ is understood to be a function of the mass $M$ and the potentials $\Phi^{(k)}_{[p-2]}$ of the solution.
The shape of the temperature function is determined by the sum of odd powers of $r_h$, ranging from $1/r_h$ when $p=2$, contributing to the curvature term, to $1/r_h^{2n+3}$ when $p=n+3$. 
An important consequence of this is that the only way to have two branches of black holes, small and large ones as for Reissner-Nordstrom-AdS black holes, is to consider a horizon of positive scalar curvature. In this case we expect phase transitions between these families. For instance a rich phase diagram for the electromagnetic case ($p=2$) with spherical horizon is well known, similar to the Van der Waals-Maxwell liquid-gas system \cite{Chamblin:1999tk}. On the other hand, black holes with flat horizons, although they can have an hyperbolic-like lapse function ($\kappa=-1$), do not undergo any phase transition.

\paragraph{Free energy:}
The free energy of these black holes can be evaluated using the path integral approach of \cite{Hartle:1976tp,Gibbons:1976ue}, in which the partition function in some thermodynamical ensemble is identified with the saddle point approximation of the Euclidean path integral, with the boundary conditions corresponding to the ensemble.
Specifically, here we will analyze the black holes in the grand canonical ensemble, in which the electric potentials of the form fields are kept constant.
The Euclidean section of the solutions with metric \eqref{metric} and \eqref{genV}, obtained by a Wick rotation, has a bolt at the largest root of $V(r)$. To avoid the associated conical singularity, and make the manifold regular, we identify the Euclidean time with period $\be=1/T$, where $T$ is the temperature, given by \eqref{genT}.

Boundary conditions, or equivalently the choice of the thermodynamical ensemble, dictate the boundary terms that must be added to the action functional in order to make the variational principle well-posed. As usual, the gravitational action must be supplemented with the Gibbons-Hawking surface term, $S_E = - S - S_{GH}$, with 
\beq
S_{GH}  = \frac{1}{8\pi G} \int_{\p \mc M} K.
\eeq
Here, $K$ is the trace of the extrinsic curvature of the boundary $\p \mc M$ of the spacetime. This surface term is necessary if we allow variations of the metric for which only the induced metric on the boundary is held fixed in order to establish the Einstein equation. On the other hand, there is no boundary term for the matter sector since this part is well defined when keeping the electric potentials fixed on the boundary. Indeed, the general variation for the matter action for a general $p$-form field reads,
\beq
\delta S_M =
-\frac{2}{(p-1)!} \int_\mc M \dd^{D}x\sqrt{-g}\,\nabla_B H^{B A_1\ldots A_{p-1}}\delta A_{A_1\ldots A_{p-1}}
+\frac{2}{(p-1)!}\int_{\p \mc M} n_B H^{B A_1\ldots A_{p-1}}
\delta A_{A_1\ldots A_{p-1}},
\eeq
where $n_A$ is the unit normal to the boundary and the natural volume element on $\p \mc  M$ is understood. Hence, this action gives a well-posed variational principle for the grand canonical ensemble.

However, the Euclidean action $S_E$ evaluated on a solution is typically not well-defined, as the integral is formally divergent. To extract its finite, physical value, we shall use the background subtraction technique, in which physical quantities are computed relatively to some reference background $\mc M_0$. We regularize the action by integrating over a finite spacetime region with boundary $\p \mc M$, and subtracting the Euclidean action evaluated on a finite region  of $\mc M_0$, chosen such that the induced metric and electric potentials on the boundary of $\mc M_0$ coincide with those on $\p\mc M$. Then, the cutoff can be safely eliminated, by taking the limit in which the boundary $\p\mc M$ is sent to spatial infinity.

Since we are working in the grand-canonical ensemble, the choice of the reference background must be such that both its temperature and its electric potentials can be matched to the respective ones of the configurations under scrutiny.
A good candidate is the Euclidean AdS vacuum with constant electric potentials $\Phi^{(k)}_{[p-2]} $. Indeed, this instanton is regular and contains no conical singularity; it can hence be assigned any  periodicity $\beta$ in Euclidean time. Then, the Gibbs potential $G$ is determined by,
\beq
\beta G = S_E - S_E^0  = \frac{\beta V_{\mc H}}{16\pi G} \left[  \kappa r_h^n - \frac{r_h^{n+2}}{\ell^2}+ \frac{1}{2(n+1)} \sum_{p,k}  (3-2p ) (n-2p+4) r_h^{n-2p+4} \frac{ \Phi^{(k)2}_{[p-2]} }{(p-2)! }   \right],
\eeq
where the expotent $^0$ denotes the background and $V_{\mc H}$ is the (possibly diverging) area of the Riemannian manifold $\mc H$. It is then straightforward to, assuming the first law of thermodynamics,
\beq
\delta M= T \delta S - \sum_{p,k} \frac{1}{(p-2)!} \Phi^{(k)}_{i_1 \ldots i_{p-2}}\delta Q_{(k)}^{i_1 \ldots i_{p-2}},
\eeq
to obtain the entropy, the brane charges and the mass of the black hole,
\eqa
&&
S =  - \left. \frac{\p G}{\p T} \right|_{ \{ \Phi^{(k)}_{[p-2]} \}_{p,k}} = \frac{V_{\mc H}}{4G}\,r_h^{n+1},\nonumber\\  
&&
Q_{(k)}^{i_1 \ldots i_{p-2}} =  (p-2)! \left. \frac{\p G}{\p \Phi^{(k)}_{i_1 \ldots i_{p-2}} } \right|_{ \left\{ T,  \Phi^{(k')}_{i'_1 \ldots i'_{p'-2}}  \neq  \Phi^{(k)}_{i_1 \ldots i_{p-2}} \right\} }   = \frac{V_{\mc H}}{16\pi G} \mc E_{(k)}^{i_1 \ldots i_{p-2}},\\
&&
M=G + TS -  \sum_{p,k} \frac{1}{(p-2)!} Q_{(k)}^{i_1 \ldots i_{p-2}}  \Phi^{(k)}_{i_1 \ldots i_{p-2}} = \frac{V_{\mc H}}{16\pi G}(n+1) r_0^n.
\eeqa
It can be readily checked that the entropy is proportional to one quarter of the horizon area in geometrized units and that the mass agrees with the expected one from an asymptotic analysis of the geometry. As for the brane-charge it is simply proportional by the electric polarization $\mc E$. We will end this section by rederiving these results using the Hamiltonian approach, putting these definitions of the charges on a firm ground.

\paragraph{Hamiltonian approach:}
In this paragraph we propose to recover the previous results by a Hamiltonian approach \cite{Hawking:1995fd,Brown:1992br} to emphasize that brane charges are quantities coming from the boundary of $\mc M$ at spatial infinity. 
The first step  in this formalism is a breakup of the spacetime $\mc M$ into space and time. We assume that there is a diffeomorphism $\phi : \mc M \rightarrow \Sigma \times I$, $I \subset \mathbb{R}$, such that the submanifolds $\Sigma_t = \phi^{-1}\left( \Sigma \times \{t\} \right)$ are spacelike and the curves $ \phi^{-1}\left(  \{x\} \times I \right)$ are timelike. A tangent vector $t^\mu$ to these curves can be decomposed as usual into $ t^\mu = N n^\mu + N^\mu$, where $n^\mu$ is the unit normal to the surface $\Sigma_t$, $N$ is the lapse function and $N^\mu$ is the shift vector. This split induces a decomposition of the boundary of $\mc M$ into an initial and a final spacelike hypersurfaces $\Sigma_{t_i}$ and $\Sigma_{t_f}$ and a timelike boundary ${}^{D-1}B$ with unit normal vector $u^\mu$. This timelike boundary will eventually be sent at spatial infinity. Moreover, we denote by $B$ the boundary of $\Sigma_t$ and we require\footnote{The reader can find in \cite{Hawking:1996ww} the case where this orthogonality condition is dropped.} $u^\mu n_\mu = 0$ on ${}^{D-1}B$. The induced metric on $\Sigma_t$ and  $B$ are $h_{\mu\nu}=g_{\mu\nu}+n_\mu n _\nu$ and $s_{\mu\nu}=h_{\mu\nu}-u_\mu u _\nu$ respectively. Then, we start with the action
\beq
S'=\alpha \int_{\mc M} \dd^D x \sqrt{-g} \left[ R - 2\La -\frac{\kappa}{p!}H^2_{[p]} \right] + 2\alpha \int_{\p \mc M} K
\eeq
which provides a well-defined variational problem where only the induced metric and the potential are fixed on the boundary $\p \mc M$. After that we follow  the standard procedure \cite{Hawking:1995fd,Wald:1984rg} to derive the Hamiltonian formulation for this theory. The momenta canonically conjugate to the spatial metric $h_{\mu\nu}$ and to the potential $\mc A_{[p-1]}$ are
\beq
\pi_G^{\mu\nu} = \alpha \sqrt{h} \left( K^{\mu\nu} - K h^{\mu\nu} \right), 
\qquad
\pi_{H}^{ \mu_2 \ldots \mu_p} = \frac{2\kappa\alpha\sqrt{h}}{(p-1)!} \left( n_{\mu_1} H^{\mu_1 \mu_2 \ldots \mu_p} \right)_{\parallel},
\eeq
where $K_{\mu\nu}=h_\mu{}^{\rho}\nabla_\rho n_\nu $ is the extrinsic curvature of $\Sigma_t$ and the symbol $\parallel$ denotes the projection operator on $\Sigma_t$. Indeed, we can view $h^\nu{}_{\mu}$ as the projection of the tangent space of $\mathcal{M}$ at $p$ on the tangent space of $\Sigma_t$ at $p$; similarly
$T_{\parallel}^{\mu_1 \ldots \mu_k}{}_{\nu_1 \ldots \nu_l} = h_{\ \rho_1}^{\mu_1} \ h_{\ \rho_k}^{\mu_k} h_{\nu_1}^{\ \sigma_1}\ldots h_{\nu_l}^{\ \sigma_l}  T^{\rho_1\ldots\rho_k}{}_{\sigma_1\ldots\sigma_l}$ for any spacetime tensor $T^{\mu_1\ldots\mu_k}{}_{\nu_1\ldots\nu_l}$. The resulting Hamiltonian is,
\begin{multline}
 H = \int_{\Sigma_t} \sqrt{h} \left (N C +  N^\mu C_\mu + t^{\mu_1} A_{\mu_1 \mu_2\ldots\mu_{p-1}} C^{\mu_2\ldots\mu_{p-1}}\right) d^{D-1}x\\
 + \int_{B} \sqrt{s} \left( -2 \alpha N k + 2 N^\mu u^\nu \frac{\pi_{G\mu\nu}}{\sqrt{h}} + (p-1) \frac{\pi_{H}^{ \mu_2\ldots\mu_p}}{\sqrt{h}} t^{\mu_1} A_{\mu_1  \mu_3\ldots\mu_p} u_{\mu_2} \right)  d^{D-2}x,
\end{multline}
where the constraints -- that have to vanish on-shell -- reduce to the field equations of the theory,
\begin{align}
 C &= \frac{1}{\alpha h}\left(\pi_{G\mu\nu} \pi_G^{\mu\nu}  + \frac{\pi_G^2}{2-D} \right)  - \alpha R^{(D-1)} + 2\alpha\Lambda + \frac{(p-1)!}{4\kappa\alpha h} \pi_{H \mu_2\ldots\mu_p} \pi_{H}^{\mu_2\ldots\mu_p} \nonumber\\
&\qquad + \frac{\kappa \alpha}{p!} H_{\parallel\mu_1\ldots\mu_p} H_{\parallel}^{ \mu_1\ldots\mu_p}
 	= - 2 \alpha \left( G_{\mu\nu} + \Lambda g_{\mu\nu} - \kappa T_{\mu\nu} \right) n^\mu n^\nu ,\\
C_\mu &= - 2 D_\nu \left( \frac{\pi_{G\mu}^\nu}{\sqrt{h}} \right)   + \frac{\pi_{H \mu_2\ldots\mu_p}}{\sqrt{h}} H_{\parallel \mu  \mu_2\ldots\mu_p}
= - 2 \alpha \left( G_{\nu\rho} + \Lambda g_{\nu\rho}  - \kappa T_{\nu\rho} \right) h_\mu^{\ \nu} n^\rho,\\
 C^{\mu_2\ldots\mu_{p-1}} & = -(p-1)D_\mu \left( \frac{\pi_H^{\mu \mu_2\ldots\mu_{p-1}}}{\sqrt{h}} \right) = - \frac{2\alpha \kappa}{(p-2)!} \left( n_{\mu_p} \nabla_{\mu_1} H^{ \mu_2\ldots\mu_p \mu_1 } \right)_{\parallel},
\end{align}
with $R^{(D-1)} $ the Ricci scalar of $\Sigma_t$. In the boundary term, $k$ denotes the trace of the extrinsic curvature of $B$ given by $k =  s_\mu{}^{\nu}D_\nu u^\mu$ with $D_\mu$ the covariant derivative on $\Sigma_t$.
We emphasize the presence of boundary terms, necessary to make the Hamiltonian a differentiable functional, and crucial to establish the first law of black hole mechanics. Here we have just considered the boundary ${}^{D-1}B$, however for black hole there is also an interior boundary due to the horizon and consequently an additional contribution in the boundary terms as we will see below. The result presented here agrees with \cite{Copsey:2005se} where the boundary terms are derived by varying the Hamiltonian.

Consequently, we can define the total energy of the solution to be the value of the on-shell Hamiltonian relative to the same background that we previously used with the path integral approach. For the action \eqref{action1}, we find
\beq
H_{\text{cl}} - H_{\text{cl}}^0 = \int_B  d^{n+1}x\,\sqrt{s} \left( \frac{-N}{8\pi G}  \left( k - k^0 \right)  + \sum_{p,k} \frac{p-1}{\sqrt{h}} u_{\mu_2} \pi_{H_{[p]}^{(k)}}^{ \mu_2\ldots\mu_p}  t^{\mu_1} A^{(k)}_{\mu_1\mu_3\ldots\mu_p}    \right),
\eeq
since it is sufficient to consider spacetimes with vanishing shift vector in our case. Moreover the static slices are labelled so that $N=N^0$ on $^{D-1}B$. After that, we require
\beq
H_{\text{cl}} - H_{\text{cl}}^0 = M + \sum_{p,k} \frac{1}{(p-2)!} Q_{(k)}^{i_1\ldots i_{p-2}}  \Phi^{(k)}_{i_1\ldots i_{p-2}},
\eeq
where $M$ defines the mass of the solution and $ Q_{(k)}^{i_1\ldots i_{p-2}} $ is the brane-charge associated to the electric potential $\Phi^{(k)}_{i_1\ldots i_{p-2}}$. In this way we can deduce the following expressions for the mass,
\beq 
M = -\frac{1}{8\pi G}  \int_B \sqrt{s} N \left( k - k^0 \right)\dd^{n+1}x,
\eeq
and for the charges,
\beq 
Q_{(k)}^{i_1\ldots i_{p-2}}  \Phi^{(k)}_{i_1\ldots i_{p-2}} = \frac{1}{16\pi G}  \int_B \sqrt{s}     \left( n_{\mu_1} H_{(k)}^{\mu_1 \mu_2\ldots \mu_p} \right)_{\parallel} u_{\mu_2}  t^\nu A^{(k)}_{\nu  \mu_3\ldots \mu_p} \dd^{n+1}x,
\eeq
of the static configurations we are interested in.
In particular, for the black holes presented in this article we find,
\beq M = \frac{V_{\mc H}}{16\pi G}(n+1) r_0^n,\qquad
Q_{(k)}^{i_1\ldots i_{p-2}}=\frac{V_{\mc H}}{16\pi G}\mc E_{(k)}^{i_1\ldots i_{p-2}}\,. 
\eeq
Thus, we obtain a mass and brane charges in agreement with those we found previously with the path integral method. However, in the Hamiltonian approach we did not need to assume the validity of the first law of thermodynamics; on the contrary, introducing the extra boundary term on the horizon, it is straightforward to show, following \cite{Wald:1993ki,Sudarsky:1992ty} that the first law of black hole mechanics,
\beq 
\delta M = \frac{\kappa_H}{8\pi}\delta A_H  - \sum_{p,k} \frac{1}{(p-2)!} \Phi_{(k)}^{i_1\ldots i_{p-2}}  \delta Q^{(k)}_{i_1\ldots i_{p-2}},
\eeq
hold, where $\kappa_H$ and $A_H$ are the surface gravity and the area of the horizon respectively.

\section{Conclusion and outlook}
\label{sec::conclusion}

In this article, we found a large number of AdS black holes dressed with free scalar/$p$-form fields, some of which survive even for $\La\geq0$. Indeed, when the horizon has positive curvature, these solutions can be continued to asymptotically locally flat/de~Sitter black holes. We did not restrict to any particular field content of the theory in our analysis, but it is important to stress that the constructions presented here can be naturally embedded in supergravity theories admitting an AdS vacuum, such as gauged supergravities, possibly arising from consistent Kaluza-Klein compactifications of $D=10$, $11$ supergravities (see for example \cite{Gauntlett:2007ma} and references therein), when the matter fields are free. Moreover, we expect these solutions to play a role in the AdS/CFT correspondence, with the matter fields deforming the dual CFT, possibly describing some condensed matter system when engineering the desired properties of the dual theory by shaping the lapse function $V(r)$, as explained in the article. We shall not comment further on this aspect, and leave the topic for future investigations.

The geometry of the $\kappa>0$ event horizons of the black holes found in Sections~\ref{sec::single} and \ref{sec::kaehler} deviates from the usual round spherical metric of the Schwarzschild-Tangherlini and its (anti-)de~Sitter generalizations. In vacuum, black holes with such horizons are known to be classically unstable \cite{gibbonshartnoll}. It would be interesting to study how the presence of the magnetic $p$-form fields affect this instability, and whether these external fields might provide a stabilization mechanism.  

We also found planar AdS black holes \eqref{magscalar} dressed with $D-2$ scalar fields, one for each direction on the horizon. In section~\ref{sec::multiple} we argued that the shift symmetry of these scalars entitles us to compactify the horizon on a $(D-2)$-torus. The resulting configuration enjoys the planar Euclidean symmetries on the horizon directions in addition to the time translation symmetry generated by $\p_t$. In other words, the change in the scalar fields induced by a isometry of the metric, say the translation generated by $\p_x$ for example, is simply a shift of the field that is pure gauge, but the physical quantities, the gradients of the scalars, remain unchanged. However, one could interpret the scalar fields differently, at the cost of giving up on the compactification to a torus, and keeping a black hole with an extended planar horizon. In this case, one can break the shift invariance of the scalars at the level of the action (for example, by coupling them linearly to some extra free scalar field that vanishes on these solutions). Then, the value assumed by the scalar field becomes a physical observable, breaking the Euclidean symmetries generated by the isometries of $\mc H$. Indeed, any of these isometries -- call $\xi$ the Killing field generating it -- while still verifying $\mc L_\xi g_{\mu\nu}=0$, changes the scalars fields since $\mc L_\xi\phi^{(i)}\neq0$, and is therefore not anymore a symmetry of the black hole configuration. The only residual symmetry is the time translation symmetry. In other words, the planar black holes dressed with scalar fields \eqref{magscalar} are very simple examples of black holes with only one Killing field, valid in any dimension, in the spirit of \cite{Dias:2011at}. Note however that, unlike the solutions presented in those works, here the horizon is not compact and the geometry is not globally asymptotically AdS.

It would be interesting to see how far one can go by relaxing the static metric ansatz \eqref{metric}. In particular, we believe it should be possible to add rotation to the four dimensional axionic black holes \eqref{tormet}-\eqref{axioniclapse}, thereby dressing the rotating cylindrical black hole of \cite{Klemm:1997ea} with two free three-form fields (or equivalently, with two free scalar fields). This dressing should extend to the case of Taub-NUT-AdS black holes, as well at to the AdS C-metric, or more generally, to the Pleba\'nski-Demia\'nski type-D geometry \cite{Plebanski:1976gy} containing them all as particular limits. Some of those, if found, might prove supersymmetric in their extremal limit, by embedding the solutions in $\mc N=2$, $D=4$ gauged supergravity coupled to abelian vector multiplets, similarly to the BPS solutions found by Klemm \cite{Klemm:2011xw}.

Also, generalizations of these solutions to more general matter content or different theories are of great interest. For example, using the techniques described here, it is possible to construct four dimensional axionic black holes of Einstein-Maxwell-AdS gravity with a conformally coupled scalar and two axionic fields, yielding regular axionic Bekenstein black holes generalizing those of \cite{mtz}. These black holes exhibit secondary hair with interesting phase transitions, and will be presented in detail in \cite{bcc}. Other possible extensions of this work deserve to be explored, such as including free spinorial fields in the matter sector, or non-trivial couplings/potentials for the $p$-form fields. While complicating the field equations, exact solutions might still be within reach, but go beyond the scope of the present article.

\section*{Acknowledgements} It is a pleasure to thank J. Camps, O. Dias, B. Gout\'eraux, R. Gregory, D. Klemm, E. Kiritsis, F. Nitti and  B. Vercnocke for many useful discussions.
This work was partially supported by the ANR grant STR-COSMO, ANR-09-BLAN-0157.
MMC was additionally partially supported by
the ERC Advanced Grant 226371,
the ITN programme PITN-GA-2009-237920,
the IFCPAR CEFIPRA programme 4104-2 and the ANR programme NT09-573739 ``string
cosmo''.


\end{document}